\documentclass[12pt]{article}
\begin{document}
\title{The effect of strain rate on the viscoplastic behavior
of isotactic polypropylene at finite strains}

\author{A.D. Drozdov and J. deC. Christiansen\\
Department of Production\\ 
Aalborg University\\
Fibigerstraede 16, DK--9220 Aalborg, Denmark}
\date{}
\maketitle

\begin{abstract}
Two series of uniaxial tensile tests are performed on isotactic 
polypropylene with the strain rates ranging from 5 to 200 mm/min.
In the first series, injection-molded specimens are used without
thermal pre-treatment.
In the other series of experiments, the samples are annealed for 51 hour
at 160 $^{\circ}$C prior to testing.

A constitutive model is developed for the viscoplastic behavior
of isotactic polypro\-pylene at finite strains.
A semicrystalline polymer is treated as an equivalent 
heterogeneous network of chains bridged by permanent junctions
(physical cross-links and entanglements).
The network is thought of as an ensemble of meso-regions connected 
with each other by links (lamellar blocks).
In the sub-yield region of deformations, junctions between chains 
in meso-domains slide with respect to their  reference positions 
(which reflects sliding of nodes in the amorphous phase 
and fine slip of lamellar blocks).
Above the yield point, the sliding process is accompanied 
by displacements of meso-domains in the ensemble with respect 
to each other (which reflects coarse slip and fragmentation of 
lamellar blocks).
To account for alignment of disintegrated lamellar blocks along
the direction of maximal stresses (which is observed as 
strain-hardening of specimens in the post-yield region of
deformations) elastic moduli are assumed to depend on 
the principal invariants of the right Cauchy--Green tensor for 
the viscoplastic flow.

Stress--strain relations for a semicrystalline polymer
are derived by using the laws of thermodynamics.
The constitutive equations are determined by 5 adjustable
parameters that are found by matching observations.
Fair agreement is demonstrated between the experimental data
and the results of numerical simulation.
A noticeable difference is revealed between the mechanical 
responses of non-annealed and annealed specimens:
(i) necking of samples not subjected to thermal treatment
precedes coarse slip and fragmentation of lamellar blocks, 
whereas cold-drawing of annealed specimens up to 
a longitudinal strain of 80\%  does not induce spatial heterogeneity 
of their deformation;
(ii) the elastic modulus increases with the strain rate
for non-annealed specimens and decreases for annealed samples.
\end{abstract}
\vspace*{5 mm}

\newpage

\section{Introduction}

This paper is concerned with the influence of strain rate on
the viscoplastic response of isotactic polypropylene (iPP)
at isothermal loading with finite strains.
The viscoplastic flow in semicrystalline polymers,
polyethylene 
\cite{BUD97,GS97,BDW98,HHL99,SL99,MP01,BKR02,Seg02,SCC02}, 
polypropylene 
\cite{AGU95,AFP95a,AFP95b,KY95,KYR95,SCC97,CCG98,SCC99,SSE99,NT99,NT00,RZF01,LVS02}
and poly(ethylene terephthalate) 
\cite{BJJ96,AGC96,MDW97,ZB97,ABJ98,LB99,SNK99,BSL00},
has been a focus of attention in the past decade, which
may be explained by the importance of the yielding 
phenomenon for the analysis of fiber- and film-drawing processes.

Injection-molded isotactic polypropylene is chosen
for the experimental analysis for two reasons:
\begin{enumerate}
\item
Polypropylene is widely used in industrial applications 
(ranged from oriented films for packaging 
to nonwoven fabrics and reinforcing fibres).

\item
Injection-moded iPP is a semicrystalline polymer with a
rather complicated crystalline morphology that is
strongly affected by annealing 
\cite{AFP95a,AFP95b,KB97,YMT98,ABM99,MHY00,IS00,GHT02}
and mechanical loading 
\cite{AGU95,CCG98,SSE99,RZF01,LVS02}.
\end{enumerate}

Isotactic polypropylene contains 
three basic crystallographic forms \cite{IS00}:
monoclinic $\alpha$ crystallites,
(pseudo) hexagonal $\beta$ structures,
orthorhombic $\gamma$ polymorphs,
and ``smectic" mesophase (arrays of chains with a better 
order in the longitudinal than in transverse chain direction).
At rapid cooling of the melt (which is typical of the 
injection-molding process), $\alpha$ crystallites 
and smectic mesophase are mainly developed,
whereas $\beta$ and $\gamma$ polymorphs are observed 
as minority components \cite{KB97}.
The characteristic size of $\alpha$ spherulites in an injection-molded 
specimen is estimated as 100 to 200 $\mu$m \cite{CCG98,KB97}.
These structures are formed by lamellae stacks 
with lamellar thicknesses ranging from 10 to 20 nm 
\cite{CCG98,MHY00}
A unique feature of the crystalline morphology of iPP is the lamellar 
cross-hatching: development of transverse lamellae oriented 
in the direction perpendicular to the direction of radial lamellae in 
spherulites \cite{IS00}.

Transmission electron microscopy \cite{YMT98,MHY00}
and polirized optical microscopy \cite{YMT98}
reveal that annealing of iPP in the sub-melting interval of
temperatures results in
\begin{itemize}
\item
a decrease in the concentration of transverse lamellae 
(and disappearance of cross-hatching after thermal treatment
above 150~$^{\circ}$C),

\item
an increase in the fraction of ordered $\alpha$ crystallites
(where helices in monoclinic unit cells are oriented in the 
same direction).
\end{itemize}

WAXS (wide-angle X-ray scattering) measurements 
\cite{MHY00,GHT02} demonstrate that annealing of 
isotactic polypropylene in 
the vicinity of a critical temperature $T_{\rm c}\approx 157$ 
to 159~$^{\circ}$C induces a second-order phase
transition in which the ordered $\alpha$ phase is replaced 
by a new crystalline form with a larger lattice volume 
and a higher mobility of chains.
As molecular mobility plays the key role in the viscoplastic
behavior of semicrystalline polymers, 
one can expect that this transformation strongly affects 
the response of iPP.
To study this phenomenon in detail, two series of uniaxial 
tensile tests are performed: 
(i) on specimens not subjected to thermal treatment, 
and 
(ii) on samples annealed for two days at 160~$^{\circ}$C
prior to loading.

Recent studies on mechanically-induced transformations of 
the crystalline structure of iPP 
\cite{AGU95,CCG98,SSE99,RZF01,LVS02}
show that loading a specimen in the sub-yield region of 
deformations results in
(i) inter-lamellar separation,
(ii) rotation and twist of lamellae,
(iii) fine slip of lamellar blocks (homogeneous shear of 
layer-like crystalline structures) \cite{GS97,HHL99,SSE99},
(iv) chain slip through the crystals \cite{NT99}, 
(v) sliding of tie chains along and their detachment from 
lamellar blocks \cite{NT99},
and (vi) activation of the rigid amorphous fraction (part of
the amorphous phase located in regions surrounded by
radial and transverse lamellae).
Straining of isotactic polypropylene in the post-yield region 
of deformations induces
(i) coarse slip of lamellar blocks 
(heterogeneous inter-lamellar sliding) and their fragmentation,
(ii) detachment of chain folds and loops from surfaces of
crystal blocks, 
(iii) disintergation of crystalline lamellae into a mosaic structure,
and (iv) alignment of broken lamellar blocks along the direction of
maximal stresses and formation of a fibrillar texture 
\cite{GS97}.

To develop a constitutive model for the mechanical response of
a semicrystalline polymer, we apply a method of homogenization.
According to this approach \cite{BKR02},
a complicated micro-structure of isotactic polypropylene is replaced
by an equivalent phase whose response captures essential features
of the mechanical behavior of the semicrystalline polymer.
An amorphous phase (treated as a network of macromolecules 
bridged by junctions) is chosen as the equivalent phase for 
the following reasons:
\begin{enumerate}
\item
The viscoplastic flow in semicrystalline polymers is 
``initiated in the amorphous phase before transitioning into the
crystalline phase" \cite{MP01}.

\item
Sliding of tie chains along and their detachment from lamellae play 
the key role in the yielding phenomenon \cite{NT99}.
\end{enumerate}

A semicrystalline polymer is thought of as an equivalent network of 
chains bridged by junctions (entanglements, physical cross-links 
on the surfaces of crystallites and lamellar blocks):
\begin{itemize}
\item
To simplify the analysis, the network is treated as incompressible
(observations show that the degree of compressibility 
of polypropylene is rather low:
the growth of the tensile strain up to 8\% causes
an increase in the volume strain by less than 1\% \cite{MF97}).
The hypothesis regarding incompressibility of semicrystalline
polymers at large deformations has been previously applied to
the analysis of the mechanical response of 
polyethylene \cite{BKR02,SCC02},
polypropylene  \cite{SCC97,SCC99}
and poly(ethylene terephthalate) \cite{BJJ96,MDW97}.

\item
To exclude from the consideration viscoelastic effects
(associated with rearrangement of chains), all nodes 
are treated as permanent (active strands cannot separate
from junctions and dangling strands cannot merge with
the network during the experimental time-scale).

\item
To describe the viscoplastic flow, the network is assumed to
deform non-affinely (junctions can slide with respect to their 
reference positions under loading).
\end{itemize}

Unlike conventional theories of non-affine networks,
two sliding processes are introduced.
The first reflects motion of nodes in amorphous
regions and fine slip of crystallites.
This process is assumed to occur at any intensity of strains.
The other is attributed to coarse slip and fragmentation of
lamellar blocks and it takes place in the post-yield region
of deformations only.
The difference between these two types of non-affine deformation 
is that sliding of junctions in amorphous domains
does not produce dissipation of energy, whereas coarse
slip and disintegration of lamellae result in a noticeable entropy
production.
At uniaxial tension, the yield point is associated with the 
elongation ratio, at which fragmentation of lamellar blocks starts.

The objective of this study is two-fold:
\begin{enumerate}
\item
To report experimental data in two series of
uniaxial tensile tests on injection-molded iPP specimens:
the samples in the first series of tests were not subjected 
to thermal treatment, whereas those in the other series of 
experiments were annealed at 160~$^{\circ}$C prior 
to loading.

\item
To develop constitutive equations for the viscoplastic behavior of
a semicrystalline polymer at finite strains
and to determine adjustable parameters in the stress--strain 
relations by fitting observations in tensile tests 
with the strain rates ranging from 
$1.5 \cdot 10^{-3}$ to $6.2 \cdot 10^{-2}$ s$^{-1}$
(practically in the entire region of cross-head speeds
employed in conventional quasi-static tests).
\end{enumerate}
Our goal is to assess the influence of thermal pre-treatment
and strain rate on the viscoplastic response of isotactic
polypropylene.

The exposition is organized as follows.
Experimental data in uniaxial tensile tests are reported in 
Section 2.
Kinematic relations for sliding of junctions 
are developed in Section 3.
Kinetic equations for the rate-of-strain tensor 
(that describes sliding of junctions in amorphous regions 
and fine slip of lamellar blocks) are introduced in Section 4.
Strain energy density of a semicrystalline polymer
is calculated in Section 5.
Constitutive equations for an isothermal deformation 
with finite strains and kinetic relations for coarse slip 
and fragmentation of lamellar blocks are derived in Section 6
by using the laws of thermodynamics.
Phenomenological expressions for material functions
are proposed in Section 7.
The governing equations are simplified for uniaxial tension of
an incompressible medium in Section 8.
Adjustable parameters in the stress--strain relations 
are determined in Section 9 by matching the 
experimental data.
The effect of strain rate on material parameters
is discussed in Section 10.
Some concluding remarks are formulated in Section 11.

\section{Experimental procedure}

Isotactic polypropylene (Novolen 1100L) was supplied by BASF (Targor).
ASTM dumbbell specimens were injection molded 
with length 148 mm, width 10 mm and thickness 3.8 mm.
Uniaxial tensile tests were performed at room temperature 
on a testing machine Instron--5568 equipped with 
electro-mechanical sensors for the control of longitudinal 
strains in the active zone of samples.
The tensile force was measured by a standard load cell.
The engineering stress, $\sigma$, was determined
as the ratio of the axial force to the cross-sectional area
of stress-free specimens.
The true longitudinal stress, $\Sigma$, was calculated 
by means of the incompressibility condition as
\[
\Sigma=\sigma k,
\]
where $k$ is the elongation ratio.

Two series of mechanical tests were carried out.
Every test was performed on a new specimen.

The first series of experiments consisted of 7 tests on 
specimens that were not subjected to thermal treatment
with the cross-head speeds 5, 10, 25, 50, 100, 150 and 200 mm/min
(which correspond to the Hencky strain rates
$1.54 \cdot 10^{-3}$,
$3.02 \cdot 10^{-3}$,
$7.22 \cdot 10^{-3}$,
$1.49 \cdot 10^{-2}$,
$3.09 \cdot 10^{-2}$,
$4.56 \cdot 10^{-2}$
and
$6.23 \cdot 10^{-2}$
s$^{-1}$, respectively).
The specimens were elongated up to the Hencky strain 
$\epsilon_{\rm H}=0.3$ that noticeably exceeded the strain 
corresponding to the onset of necking at all cross-head speeds.
The chosen strain rates ensured nearly isothermal 
experimental conditions, on the one hand,
and they allowed the viscoelastic effects to be neglected,
on the other.
The duration of tensile tests before necking of samples
varied from 4 s for the highest cross-head speed to 180 s 
for the lowest one.
Our recent study demonstrated that the amount of stress
relaxing during this period did not exceed 10\% \cite{DC02a}.

The other series of experiments was performed on specimens 
that were annealed in an oven for 51 h at 160 $^{\circ}$C 
and cooled by air.
To minimize the effect of physical aging, mechanical tests were 
carried out one day after thermal treatment.
The series consisted of 7 tests that were carried out with the same 
cross-head speeds as the experiments on non-annealed samples.
The specimens were stretched up to the maximal Hencky strain, 
$\epsilon_{\rm H}=0.6$.
No necking of samples was observed in all experiments, 
except for the test with the maximal strain rate, 
$\dot{\epsilon}_{\rm H}=6.23 \cdot 10^{-2}$ s$^{-1}$,
where the onset of a weak neck was revealed at the elongation ratio 
$k\approx 1.75$.

The true longitudinal stress, $\Sigma$, is plotted versus 
the elongation ratio, $k$, in Figures 1 to 7 for the non-annealed specimens
(only the experimental data below the necking points are presented)
and in Figures 8 to 14 for the samples subjected to thermal
pre-treatment.

A noticeable difference is observed between the engineering 
stress--engineering strain curves for non-annealed and annealed
specimens.

A typical stress--strain diagram for a sample that was not subjected
to thermal treatment demonstrates an increase in the engineering
stress with the engineering strain below some strain $\epsilon_{\rm y}$
(close to the yield strain, $\epsilon_{\rm y}^{\circ}=0.13$, provided by
the supplier) and a very weak decrease (practically a horizontal
plateau) in an interval between $\epsilon_{\rm y}$ 
and a strain, $\epsilon_{\rm n}$, at which a neck is formed.
The presence of a rather wide region of maxima on the engineering 
stress--engineering strain curve does not allow the yield stress
and the yield strain for a non-annealed specimen to be determined
with a high level of accuracy
(in agreement with the conventional approach in viscoplasticity
of solid polymers \cite{QPR97}, the yield point is associated here with 
the point of maximum on the engineering stress--engineering strain 
diagram).
This situation is rather typical for semicrystalline polymers
that are not subjected to thermal treatment prior to mechanical
testing: a similar observation was reported in \cite{BJJ96} for
poly(ethylene terephthalate).

A stress--strain curve for an annealed sample reveals
an increase in the engineering stress below the yield points,
a sharp maximum on the engineering stress--engineering strain
diagram, and a monotonical decrease in the engineering stress 
in the post-yield region of deformations.
The true stresses, $\Sigma_{\rm y}$, corresponding to the points
of maxima on the engineering stress--engineering strain curves,
as well as the appropriate elongation ratios, $k_{\rm y}$, 
are plotted in Figure 15 versus the Hencky strain rate, 
$\dot{\epsilon}_{\rm H}$.
This figure shows that the true yield stress, $\Sigma_{\rm y}$,
monotonically increases with $\dot{\epsilon}_{\rm H}$,
while the elongation ratio for yielding, $k_{\rm y}$, is practically
independent of the strain rate.
It is worth noting that the yield strains for non-annealed specimens,
$\epsilon_{\rm y}$, exceed the engineering yield strains, 
$\epsilon_{\rm y}=k_{\rm y}-1$, for annealed specimens by a factor 
of 2 to 3.

The experimental data depicted in Figure 15 are fitted
by the conventional equation (that can be developed within
the concept of thermal activation of screw dislocations \cite{BDW98})
\begin{equation}
\Sigma_{\rm y}=
\Sigma_{0}+\Sigma_{1}\log \dot{\epsilon}_{\rm H},
\end{equation}
where $\log=\log_{10}$ and the adjustable parameters $\Sigma_{m}$
($m=0,1$) are determined by the least-squares technique.
Figure 15 demonstrates fair agreement between the observations
and their approximation by Eq. (1).
It should, however, be noted that the area of applicability of the 
phenomenological relation (1) is rather narrow: it cannot be employed
at very low strain rates (when the viscoelastic effects become important) 
and at high rates of straining (when rate-induced fracture mechanisms 
are dominant).

Our purpose now is to derive constitutive equations for the viscoplastic
behavior of a semicrystalline polymer that can correctly match
the experimental data plotted in Figures 1 to 14.

\section{Kinematics of sliding}

Denote by ${\bf r}_{0}$ the radius vector of an arbitrary point
in the initial state (before external loads are applied)
and by ${\bf r}(t)$ its radius vector in
the deformed state at time $t\geq 0$.
Transition from the initial state of a network to its actual 
state is determined by the deformation gradient
\begin{equation}
{\bf F}(t)=\frac{\partial {\bf r}(t)}{\partial {\bf r}_{0}}.
\end{equation}
Sliding of junctions in the network with respect to their initial
positions is treated as a transformation of the reference state, 
when a point with the initial radius vector ${\bf r}_{0}$ moves 
to the point with a radius vector ${\bf r}_{\rm s}(t)$.
This transformation is described by the deformation gradient
\[
{\bf F}_{\rm s}(t)=\frac{\partial {\bf r}_{\bf s}(t)}{\partial {\bf r}_{0}}.
\]
It is worth noting that a smooth mapping, 
${\bf r}_{\rm s}(t,{\bf r}_{0})$, is determined only locally,
which means that the new (intermidiate) configuration is not necessary
Euclidean \cite{Hau00}.

A semicrystalline polymer is modelled as an incompressible 
medium, which implies that its deformation from the reference 
state to the deformed state is isochoric (volume-preserving).
We suppose that transformation of the initial reference
state into the new reference state
(characterized by the deformation gradient ${\bf F}_{\rm s}$)
is volume-preserving as well.

Transformation of the new reference state into the deformed state
is determined by the deformation gradient
\[
{\bf F}_{\rm e}(t)=\frac{\partial {\bf r}(t)}{\partial {\bf r}_{\rm s}(t)}.
\]
The subscript indices ``s" and ``e" indicate that appropriate
deformation gradients describe sliding of junctions and elastic 
deformation (the latter means that the strain energy density is a
function of ${\bf F}_{\rm e}$).

According to the chain rule for differentiation, the tensors ${\bf F}(t)$,
${\bf F}_{\rm s}(t)$ and ${\bf F}_{\rm e}(t)$ are connected by the
relationship
\begin{equation}
{\bf F}(t)={\bf F}_{\rm e}(t)\cdot {\bf F}_{\rm s}(t),
\end{equation}
where the dot stands for inner product.
Formula (3) coincides with the multiplicative presentation of the
deformation gradient proposed in \cite{BJJ96}, 
where ${\bf F}_{\rm e}$ is called the ``network stretch" tensor 
and ${\bf F}_{\rm s}$ is referred to as the ``slippage stretch" tensor.

Differentiation of Eq. (2) with respect to time implies that
\[
\frac{d{\bf F}}{dt}(t)=\frac{\partial {\bf v}(t)}{\partial {\bf r}(t)}
\cdot \frac{\partial {\bf r}(t)}{\partial {\bf r}_{0}},
\]
where ${\bf v}(t)=d{\bf r}(t)/dt$ is the velocity vector.
Introducing the velocity gradient
\[ {\bf L}(t)=\frac{\partial {\bf v}(t)}{\partial {\bf r}(t)}, \]
and using Eq. (2), we obtain
\begin{equation}
\frac{d{\bf F}}{dt}(t)={\bf L}(t)\cdot {\bf F}(t).
\end{equation}
Bearing in mind that
\[ 
\frac{d{\bf F}^{-1}}{dt}(t) =- {\bf F}^{-1}(t) \cdot 
\frac{d{\bf F}}{dt}(t)\cdot {\bf F}^{-1}(t), 
\]
we find from Eq. (4) that
\begin{equation}
\frac{d{\bf F}^{-1}}{dt} (t) =-{\bf F}^{-1}(t) \cdot {\bf L}(t).
\end{equation}
By analogy with Eqs. (4) and (5), we write
\begin{equation}
\frac{d{\bf F}_{\rm s}}{dt}(t)={\bf L}_{\rm s}(t)\cdot {\bf F}_{\rm s}(t),
\qquad
\frac{d{\bf F}_{\rm s}^{-1}}{dt}(t)=-{\bf F}_{\rm s}^{-1}(t)\cdot {\bf L}_{\rm s}(t),
\end{equation}
where 
\[
{\bf L}_{\rm s}(t)=\frac{\partial {\bf v}_{\rm s}(t)}{\partial {\bf r}_{\rm s}(t)}
\]
is the velocity gradient for sliding of junctions, and
${\bf v}_{\rm s}(t)=d{\bf r}_{\rm s}(t)/dt$.

It follows from Eq. (3) that
\[
\frac{d{\bf F}_{\rm e}}{dt}(t)
=\frac{d}{dt} \Bigl [ {\bf F}(t)\cdot {\bf F}_{\rm s}^{-1}(t) \Bigr ]
=\frac{d{\bf F}}{dt}(t)\cdot {\bf F}_{\rm s}^{-1}(t)
+{\bf F}(t)\cdot \frac{d{\bf F}_{\rm s}^{-1}}{dt}(t).
\]
Substitution of Eqs. (4) and (6) into this equality results in
\begin{equation}
\frac{d{\bf F}_{\rm e}}{dt}(t)={\bf L}(t)\cdot {\bf F}_{\rm e}(t)
-{\bf F}_{\rm e}(t)\cdot  {\bf L}_{\rm s}(t).
\end{equation}
The left and right Cauchy--Green tensors for elastic deformation are given by
\begin{equation}
{\bf B}_{\rm e}(t)={\bf F}_{\rm e}(t)\cdot {\bf F}_{\rm e}^{\top}(t),
\qquad
{\bf C}_{\rm e}(t)={\bf F}_{\rm e}^{\top}(t)\cdot {\bf F}_{\rm e}(t),
\end{equation}
where $\top$ stands for transpose.
We differentiate the second equality in Eq. (8) with respect to time, 
use Eq. (7), and find that
\begin{equation}
\frac{d{\bf C}_{\rm e}}{dt}(t)=2{\bf F}_{\rm e}^{\top}(t)\cdot {\bf D}(t)\cdot {\bf F}_{\rm e}(t)
-{\bf L}_{\rm s}^{\top}(t)\cdot {\bf C}_{\rm e}(t)
-{\bf C}_{\rm e}(t)\cdot {\bf L}_{\rm s}(t),
\end{equation}
where
\begin{equation}
{\bf D}(t)=\frac{1}{2}\Bigl [ {\bf L}(t)+{\bf L}^{\top}(t) \Bigr ] 
\end{equation}
is the rate-of-strain tensor.
Taking into account that
\[
\frac{d{\bf C}_{\rm e}^{-1}}{dt} (t) =- {\bf C}_{\rm e}^{-1}(t) \cdot 
\frac{d{\bf C}_{\rm e}}{dt}(t) \cdot {\bf C}_{\rm e}^{-1}(t),
\]
and using Eq. (9), we arrive at the formula
\begin{equation}
\frac{d{\bf C}_{\rm e}^{-1}}{dt} (t) 
=-2 {\bf F}_{\rm e}^{-1}(t) \cdot {\bf D}(t)\cdot \Bigl [ {\bf F}_{\rm e}^{-1}(t)\Bigr ]^{\top} 
+{\bf C}_{\rm e}^{-1}(t)\cdot {\bf L}_{\rm s}^{\top}(t)
+{\bf L}_{\rm s}(t)\cdot {\bf C}_{\rm e}^{-1}(t).
\end{equation}
The first principal invariant of the right Cauchy--Green tensor, 
${\bf C}_{\rm e}(t)$, reads
\[
J_{1}(t)={\bf C}_{\rm e}(t):{\bf I},
\]
where ${\bf I}$ is the unit tensor and the colon stands for convolution.
Differentiating this equality with respect to time and employing
Eqs. (8) and (9), we obtain
\begin{equation}
\frac{dJ_{1}}{dt}(t)=2\Bigl [ {\bf B}_{\rm e}(t):{\bf D}(t)
-{\bf C}_{\rm e}(t):{\bf D}_{\rm s}(t)\Bigr ],
\end{equation}
where
\begin{equation}
{\bf D}_{\rm s}(t)=\frac{1}{2}\Bigl [ {\bf L}_{\rm s}(t)
+{\bf L}_{\rm s}^{\top}(t) \Bigr ] 
\end{equation}
is the rate-of-strain tensor for sliding of junctions.
For an incompressible medium, the second principal invariant of
the right Cauchy--Green tensor, ${\bf C}_{\rm e}(t)$, is given by
\[
J_{2}(t)={\bf C}_{\rm e}^{-1}(t):{\bf I}.
\]
Differentiation of this equality with respect to time and
use of Eqs. (8) and (11) imply that
\begin{equation}
\frac{dJ_{2}}{dt}(t)=-2 \Bigl [ {\bf B}_{\rm e}^{-1}(t):{\bf D}(t)
-{\bf C}_{\rm e}^{-1}(t):{\bf D}_{\rm s}(t) \Bigr ].
\end{equation}
It follows from Eqs. (12) and (14) that the derivative of an arbitrary 
smooth function, $\Phi(J_{1},J_{2})$, of the first two principal invariants 
of the right Cauchy--Green tensor, ${\bf C}_{\rm e}(t)$,
is determined by the formula
\begin{eqnarray}
\frac{d\Phi}{dt}\Bigl (J_{1}(t),J_{2}(t)\Bigr ) &=& 
2\biggl \{ \Bigl [ \Phi_{1}(t){\bf B}_{\rm e}(t)
-\Phi_{2}(t) {\bf B}_{\rm e}^{-1}(t) \Bigr ]:{\bf D}(t)
\nonumber\\
&&-\Bigl [ \Phi_{1}(t){\bf C}_{\rm e}(t)
-\Phi_{2}(t) {\bf C}_{\rm e}^{-1}(t) \Bigr ]:{\bf D}_{\rm s}(t) \biggr \},
\end{eqnarray}
where
\begin{equation}
\Phi_{m}(t)=\frac{\partial \Phi}{\partial J_{m}}\Bigl (J_{1}(t),J_{2}(t)\Bigr )
\qquad
(m=1,2).
\end{equation}
To describe strain-hardening of a semicrystalline polymer under active
loading in the post-yield region of deformations, 
we introduce the deformation gradient 
\begin{equation}
{\bf f}(t)={\bf F}_{\rm s}(t)\cdot {\bf F}_{\rm s}^{-1}(t_{\rm y})
\end{equation}
from the reference state at the yielding point to the reference state 
at time $t\geq t_{\rm y}$, where $t_{\rm y}$ is the instant when yielding occurs.
It follows from Eqs. (6) and (17) that the tensor ${\bf f}$ 
obeys the linear differential equation
\begin{equation}
\frac{d{\bf f}}{dt}(t)={\bf L}_{\rm s}(t)\cdot {\bf f}(t),
\qquad
{\bf f}(t_{\rm y})={\bf I}.
\end{equation}
By analogy with Eq. (8), we introduce the left and right 
Cauchy--Green tensors for the post-yield transformation of
the reference state
\begin{equation}
{\bf b}(t)={\bf f}(t)\cdot {\bf f}^{\top}(t),
\qquad
{\bf c}(t)={\bf f}^{\top}(t)\cdot {\bf f}(t).
\end{equation}
Differentiating the second equality in Eq. (19) and using Eqs. (17)
and (18), we obtain
\begin{equation}
\frac{d{\bf c}}{dt}(t)=2{\bf f}^{\top}(t)\cdot {\bf D}_{\rm s}(t)\cdot {\bf f}(t).
\end{equation}
Taking into account that
\[ 
\frac{d{\bf c}^{-1}}{dt}(t)=-{\bf c}^{-1}(t)\cdot \frac{d{\bf c}}{dt}(t)\cdot {\bf c}^{-1}(t),
\]
we find from Eq. (20) that
\begin{equation}
\frac{d{\bf c}^{-1}}{dt}(t)=-2{\bf f}^{-1}(t)\cdot {\bf D}_{\rm s}(t)
\cdot \Bigl [{\bf f}^{-1}(t)\Bigr ]^{\top}.
\end{equation}
As any transformation of the reference state is volume preserving,
the first two principal invariants of the tensor ${\bf c}(t)$ are given by
\begin{equation}
j_{1}(t)={\bf c}(t):{\bf I},
\qquad
j_{2}(t)={\bf c}^{-1}(t):{\bf I}.
\end{equation}
Differentiating Eqs. (22) with respect to time and using Eqs. (19) to (21), 
we arrive at the formulas
\begin{equation}
\frac{dj_{1}}{dt}(t)=2 {\bf b}(t):{\bf D}_{\rm s}(t),
\qquad
\frac{dj_{2}}{dt}(t)=-2 {\bf b}^{-1}(t):{\bf D}_{\rm s}(t).
\end{equation}
It follows from Eq. (23) that the derivative of an arbitrary 
smooth function, $\phi(j_{1},j_{2})$, of the principal invariants 
of the right Cauchy--Green tensor, ${\bf c}(t)$, reads
\begin{equation}
\frac{d\phi}{dt}\Bigl (j_{1}(t),j_{2}(t)\Bigr ) =
2\Bigl [ \phi_{1}(t){\bf b}(t)-\phi_{2}(t){\bf b}^{-1}(t) \Bigr ]: {\bf D}_{\rm s}(t),
\end{equation}
where
\begin{equation}
\phi_{m}(t)=\frac{\partial \phi}{\partial j_{m}}\Bigl (j_{1}(t),j_{2}(t)\Bigr )
\qquad
(m=1,2).
\end{equation}
Equations (19), (22) and (24) are valid for any $t\geq t_{\rm y}$.
In the sub-yield region of deformations, when $t<t_{\rm y}$,
we set formally
\begin{equation}
{\bf b}(t)={\bf c}(t)={\bf I}, \quad
j_{1}(t)=j_{2}(t)=3,
\quad
\frac{d\phi}{dt}\Bigl (j_{1}(t),j_{2}(t)\Bigr ) =0.
\end{equation}

Our aim now is to express the rate-of-strain tensor for sliding 
of junctions, ${\bf D}_{\rm s}(t)$, in terms of the rate-of-strain tensor 
for macro-deformation, ${\bf D}(t)$, and some tensors 
that characterize elastic deformation of a specimen.

\section{Kinetics of fine slip}

A semicrystalline polymer is treated as a strongly heterogeneous
network of chains linked by junctions.
Unlike \cite{SCC99}, where the spatial inhomogeneity of the network
is associated with micronecking driven by fragmentation
of lamellar blocks, we attribute the heterogeneity of the network 
to an inhomogeneity of interactions between chains 
in the amorphous phase and crystalline lamellae with 
various lengths and thicknesses.

The network is thought of as an ensemble of meso-regions 
with arbitrary shapes and sizes.
The characteristic length of a MR substantially exceeds the radius of
gyration for a macromolecule, and it is noticeably less than the characteristic
size of a sample.

Deformation of a specimen induces two sliding processes in the network:
\begin{enumerate}
\item
Sliding of junctions between chains with respect to their 
reference positions in stress-free meso-domains.

\item
Sliding of MRs in the ensemble with respect to each other.
\end{enumerate}
Sliding of nodes in meso-domains of an equivalent network
reflects
\begin{itemize}
\item 
sliding of junctions between chains in the amorphous phase, 

\item
slippage of tie chains along lamellar surfaces \cite{NT99},

\item
fine slip of crystalline lamellae (homogeneous shearing of 
layer-like crystallites and small displacements of lamellar 
blocks with respect to one another) \cite{HHL99}.
\end{itemize}
Sliding of MRs with respect to each other describes
\begin{itemize}
\item
coarse slip of lamellae (inter-lamellar sliding), 

\item
fragmentation of lamellae into mosaic blocks linked by tie chains, 

\item
their alignment along the direction of maximal stresses 
\cite{HHL99}.
\end{itemize}

To describe evolution of the microstructure of a semicrystalline
polymer driven by simultaneous effects of these two sliding
processes, we introduce two rate-of-strain tensors, ${\bf D}_{\rm f}$ 
and ${\bf D}_{\rm c}$, and adopt the conventional assumption that
the total rate-of-strain tensor, ${\bf D}_{\rm s}$, equals the
sum of ${\bf D}_{\rm f}$ and ${\bf D}_{\rm c}$,
\begin{equation}
{\bf D}_{\rm s}={\bf D}_{\rm f}+{\bf D}_{\rm c},
\end{equation}
where the subscript indices ``f" and ``c" refer to fine and coarse
slip, respectively.

We suppose that deformation of a specimen induces sliding 
of junctions in amorphous regions and fine slip of lamellar blocks
both in the sub-yield and post-yield regions of deformation.
Te rate-of-strain tensor, ${\bf D}_{\rm f}$, is assumed to be
proportional to the rate-of-strain tensor, ${\bf D}$, 
\begin{equation}
{\bf D}_{\rm f}(t)=\alpha (t) {\bf D}(t).
\end{equation}
For an incompressible isotropic medium, 
the coefficient of proportionality, $\alpha$, depends on
the principal invariants, $J_{1}$ and $J_{2}$, of
the right Cauchy--Green tensor ${\bf C}_{\rm e}$.
This coefficient vanishes at the zero elastic strain, 
monotonically increases with elastic deformation,
and tends to some constant $a\in [0,1]$ (the rate of 
sliding of junctions for a developed viscoplastic flow) 
at relatively large elastic strains.
The inequality $a\geq 0$ means that junctions move 
in the direction determined by the macro-strain,
whereas the condition $a\leq 1$ ensures that the rate of 
the steady flow of nodes does not exceed the rate of macro-strain.

Assuming the tensor ${\bf D}_{\rm c}(t)$ to vanish in the sub-yield 
region of deformations and substituting Eqs. (27) and (28) 
into Eqs. (15) and (26), we find that for any $t<t_{\rm y}$,
\begin{eqnarray}
\frac{d\Phi}{dt}\Bigl (J_{1}(t),J_{2}(t)\Bigr ) &=& 
2\Bigl [ \Phi_{1}(t)\Bigl ( {\bf B}_{\rm e}(t)-\alpha(t){\bf C}_{\rm e}(t)\Bigr )
\nonumber\\
&&-\Phi_{2}(t) \Bigl ( {\bf B}_{\rm e}^{-1}(t) -\alpha (t){\bf C}_{\rm e}^{-1}(t) \Bigr )
\Bigr ]:{\bf D}(t),
\nonumber\\
\frac{d\phi}{dt}\Bigl (j_{1}(t),j_{2}(t)\Bigr ) &=& 0.
\end{eqnarray}
It follows from Eqs. (15), (24), (27) and (28) that
in the post-yield region of deformations, when $t\geq t_{\rm y}$, 
the derivatives of the functions $\Phi$ and $\phi$ read
\begin{eqnarray}
\frac{d\Phi}{dt}\Bigl (J_{1}(t),J_{2}(t)\Bigr ) &=& 
2\biggl \{ \Bigl [ \Phi_{1}(t)\Bigl ( {\bf B}_{\rm e}(t)-\alpha(t){\bf C}_{\rm e}(t)\Bigr )
\nonumber\\
&&-\Phi_{2}(t) \Bigl ( {\bf B}_{\rm e}^{-1}(t) -\alpha (t){\bf C}_{\rm e}^{-1}(t) \Bigr )
\Bigr ]:{\bf D}(t)
\nonumber\\
&&-\Bigl [ \Phi_{1}(t){\bf C}_{\rm e}(t)
-\Phi_{2}(t) {\bf C}_{\rm e}^{-1}(t) \Bigr ]:{\bf D}_{\rm c}(t) \biggr \},
\nonumber\\
\frac{d\phi}{dt}\Bigl (j_{1}(t),j_{2}(t)\Bigr ) &=&
2\alpha(t) \Bigl [ \phi_{1}(t){\bf b}(t)-\phi_{2}(t){\bf b}^{-1}(t) \Bigr ]: {\bf D}(t)
\nonumber\\
&&+2 \Bigl [ \phi_{1}(t){\bf b}(t)-\phi_{2}(t){\bf b}^{-1}(t) \Bigr ]: {\bf D}_{\rm c}(t),
\end{eqnarray}
where the functions $\Phi_{m}$ and $\phi_{m}$ ($m=1,2$) are 
given by Eqs. (16) and (25).

\section{Strain energy density}

We assume that the macro-strain is transmitted to individual 
meso-regions by links  (crystalline lamellae) that 
connect MRs in an ensemble.
A meso-region is treated as an incompressible isotropic medium 
with the strain energy density (per unit volume)
\[
\tilde{w} =\mu w ,
\]
where $\mu$ is an average rigidity of a meso-domain
and $w=w(J_{1},J_{2})$ is a dimensionless function that
satisfies the condition
\[
w(J_{1},J_{2})\biggl |_{J_{1}=3,\,J_{2}=3}=0.
\]
This equality means that the mechanical energy of a MR vanishes 
in the reference state.

It is postulated that the rigidity, $\mu$, is constant 
in the sub-yield region, and it monotonically
increases with viscoplastic strains in the post-yield region
of deformations, where $\mu$ becomes a function of 
the principal invariants of the right Cauchy--Green 
tensor ${\bf c}$: $\mu=\mu(j_{1},j_{2})$.
This dependence reflects strain-hardening of a semicrystalline 
polymer induced by texture formation.

We adopt the conventional assumptions that (i) the excluded-volume
effect and other multi-chain effects are screened for an
individual chain in a network by surrounding macromolecules, 
and (ii) the energy of interaction between chains in a meso-region
and between meso-domains can be taken into account 
with the help of the incompressibility condition.
These hypotheses imply that the strain energy density (per unit 
volume) of a network, $W$, equals the sum of the mechanical 
energies of MRs,
\begin{equation}
W(t)=M \Bigl (j_{1}(t),j_{2}(t)\Bigr ) w\Bigl (J_{1}(t),J_{2}(t)\Bigr ),
\end{equation}
where $M=\mu N$ is the rigidity of the network, 
and $N$ is the average number of MRs per unit volume.

Differentiating Eq. (31) with respect to time, using Eq. (29),
and taking into account that the first principal invariant of the rate-of-strain
tensor, ${\bf D}$, valishes for isochoric deformations,
we find that in the sub-yield region of deformations, when $t<t_{\rm y}$,
\begin{equation}
\frac{dW}{dt}(t) = {\bf A}^{\prime}(t):{\bf D}(t),
\end{equation}
where the prime stands for the deviatoric component of a tensor, 
\begin{equation}
{\bf A}(t) =2 M_{0}(0) \Bigl [ w_{1}(t)\Bigl ( {\bf B}_{\rm e}(t)-\alpha(t){\bf C}_{\rm e}(t)\Bigr )
-w_{2}(t) \Bigl ( {\bf B}_{\rm e}^{-1}(t) -\alpha (t){\bf C}_{\rm e}^{-1}(t) \Bigr ) \Bigr ] .
\end{equation}
Bearing in mind that the first principal invariants of 
the rate-of-strain tensors, ${\bf D}$ and ${\bf D}_{\rm c}$, 
equal zero for a volume-preserving transformation,
we find from Eqs. (30) and (31) 
that in the post-yield region of deformations, when $t\geq t_{\rm y}$,
\begin{equation}
\frac{dW}{dt}(t) = {\bf A}^{\prime}(t):{\bf D}(t)-{\bf A}_{\rm c}^{\prime}(t):{\bf D}_{\rm c}(t),
\end{equation}
where
\begin{eqnarray}
\hspace*{-5 mm}
{\bf A}(t) &=& 2 \biggl \{ M_{0}(t)\Bigl [ w_{1}(t)\Bigl ( 
{\bf B}_{\rm e}(t)-\alpha(t){\bf C}_{\rm e}(t)\Bigr )
-w_{2}(t) \Bigl ( {\bf B}_{\rm e}^{-1}(t) -\alpha (t){\bf C}_{\rm e}^{-1}(t) \Bigr ) \Bigr ]
\nonumber\\
&&+\alpha(t)w_{0}(t) \Bigl [ M_{1}(t){\bf b}(t)-M_{2}(t){\bf b}^{-1}(t) \Bigr ] \biggr \},
\nonumber\\
\hspace*{-5 mm}
{\bf A}_{\rm c}(t) &=& 2 \biggl \{ M_{0}(t) \Bigl [ w_{1}(t){\bf C}_{\rm e}(t)
-w_{2}(t) {\bf C}_{\rm e}^{-1}(t) \Bigr ]
-w_{0}(t) \Bigl [ M_{1}(t){\bf b}(t)-M_{2}(t){\bf b}^{-1}(t) \Bigr ] \biggr \}.
\end{eqnarray}
The functions $w_{m}(t)$ and $M_{m}(t)$ ($m=0,1,2$) in Eqs. (33) and (35) read
\begin{eqnarray}
w_{0}(t) &=& w\Bigl (J_{1}(t),J_{2}(t)\Bigr ),
\qquad
w_{m}(t)=\frac{\partial w}{\partial J_{m}}\Bigl (J_{1}(t),J_{2}(t)\Bigr ),
\nonumber\\
M_{0}(t) &=& M\Bigl (j_{1}(t),j_{2}(t)\Bigr ),
\qquad
M_{m}(t) = \frac{\partial M}{\partial j_{m}}\Bigl (j_{1}(t),j_{2}(t)\Bigr ).
\end{eqnarray}
Our purpose now is to derive stress--strain relations for a semicrystalline
polymer and kinetic equations for the evolution of the rate-of-strain 
tensor, ${\bf D}_{\rm c}(t)$, in the post-yield region of deformations
by using the laws of thermodynamics.

\section{Constitutive equations}

For isothermal deformation of an incompressible medium at a reference
temperature $T_{0}$, the Clausius-Duhem inequality reads \cite{Hau00}
\begin{equation}
T_{0}\frac{dQ}{dt}(t)=-\frac{dW}{dt}(t)+{\bf \Sigma}^{\prime}(t):{\bf D}(t) \geq 0,
\end{equation}
where ${\bf \Sigma}$ is the Cauchy stress tensor,
and $Q$ is the entropy production per unit volume.

Substition of Eqs. (32) and (34) into Eq. (37) implies that in the sub-yield
region of deformations, when $t<t_{\rm y}$,
\begin{equation}
T_{0}\frac{dQ}{dt}(t) = \Bigl [{\bf \Sigma}(t)-{\bf A}(t)\Bigr ]^{\prime}:{\bf D}(t),
\end{equation}
whereas 
in the post-yield region, when $t\geq t_{\rm y}$,
\begin{equation}
T_{0}\frac{dQ}{dt}(t)=\Bigl [{\bf \Sigma}(t)-{\bf A}(t)\Bigr ]^{\prime}:{\bf D}(t)
+{\bf A}_{\rm c}^{\prime}(t):{\bf D}_{\rm c}(t).
\end{equation}
The main hypothesis of the present study is that 
\begin{itemize}
\item 
sliding of nodes in the amorphous phase and fine slip of lamellar 
blocks do not induce dissipation of mechanical energy, 

\item
an increase in the specific entropy of an equivalent network is driven
by coarse slip and fragmentation of lamellae only.
\end{itemize}
This assumption implies that in the sub-yield region of deformations,
\begin{equation}
\frac{dQ}{dt}(t)=0.
\end{equation}
The entropy production in the post-yield region of deformations
is attributed to the coarse slip of lamellar blocks 
characterized by the rate-of-strain tensor ${\bf D}_{\rm c}$.
Adopting the conventional formula for the rate of entropy production
driven by a viscoplastic flow, we find that
\begin{equation}
\frac{dQ}{dt}(t)=\frac{1}{\gamma T_{0}D_{\rm i}(t)} {\bf D}_{\rm c}(t): {\bf D}_{\rm c}(t),
\end{equation}
where $\gamma$ is the rate of viscoplastic flow and
\begin{equation}
D_{\rm i}=\Bigl (\frac{2}{3}{\bf D}:{\bf D}\Bigr )^{\frac{1}{2}}
\end{equation}
is the intensity of macro-strain rate.

Substitution of expression (40) into Eq. (38) implies that for an
arbitrary loading program, the Clausius--Duhem inequality 
is satisfied in the sub-yield region of deformations, provided
that the Cauchy stress tensor, ${\bf \Sigma}$, reads
\begin{eqnarray}
{\bf \Sigma}(t) &=& -p(t){\bf I}+2 M_{0}(0) \Bigl [ w_{1}(t)
\Bigl ( {\bf B}_{\rm e}(t)-\alpha(t){\bf C}_{\rm e}(t)\Bigr )
\nonumber\\
&& -w_{2}(t) \Bigl ( {\bf B}_{\rm e}^{-1}(t) -\alpha (t){\bf C}_{\rm e}^{-1}(t) \Bigr ) \Bigr ],
\end{eqnarray}
where $p(t)$ is pressure.

It follows from Eqs. (39) and (41) that for an arbitrary 
loading program, the Clausius--Duhem inequality 
is satisfied in the post-yield region of deformations
provided that the Cauchy stress tensor, ${\bf \Sigma}$, 
is given by
\begin{eqnarray}
{\bf \Sigma}(t) &=& -p(t){\bf I}+ 
2 \biggl \{ M_{0}(t)\Bigl [ w_{1}(t)\Bigl ( {\bf B}_{\rm e}(t)-\alpha(t){\bf C}_{\rm e}(t)\Bigr )
\nonumber\\
&& -w_{2}(t) \Bigl ( {\bf B}_{\rm e}^{-1}(t) -\alpha (t){\bf C}_{\rm e}^{-1}(t) \Bigr ) \Bigr ]
\nonumber\\
&&+\alpha(t)w_{0}(t) \Bigl [ M_{1}(t){\bf b}(t)-M_{2}(t){\bf b}^{-1}(t) \Bigr ] \biggr \}
\end{eqnarray}
and the rate-of-strain tensor for coarse slip of lamellar blocks
is determined by the formula
\begin{eqnarray}
{\bf D}_{\rm c}(t) &=& 2 \gamma \biggl \{ M_{0}(t) \Bigl [ w_{1}(t){\bf C}_{\rm e}(t)
-w_{2}(t) {\bf C}_{\rm e}^{-1}(t) \Bigr ]
\nonumber\\
&& -w_{0}(t) \Bigl [ M_{1}(t){\bf b}(t)-M_{2}(t){\bf b}^{-1}(t) \Bigr ] \biggr \}^{\prime}
D_{\rm i}(t).
\end{eqnarray}
Equation (45) implies that the rate-of-strain tensor for viscoplastic flow in
the post-yield region, ${\bf D}_{\rm c}$, is proportional to the intensity
of the rate-of-strain tensor for macro-deformation, ${\bf D}$.

The set of constitutive equations for a semicrystalline polymer
consists of the stress--strain relations (43) and (44) 
together with the kinetic equations (27), (28) and (45) 
for fine and coarse slip of lamellar blocks.
The governing equations are determined by 3 material functions: 
$\alpha(t)$, $M(j_{1},j_{2})$ and $w(J_{1},J_{2})$.

For an arbitrary three-dimensional deformation,
the set of governing equations should be completed by
a relationship between the vorticity tensor for the viscoplastic flow,
${\bf \Omega}_{s}(t)=\frac{1}{2} [ {\bf L}_{s}^{\top}-{\bf L}_{s}(t) ]$,
and the vorticity tensor for macro-deformation,
${\bf \Omega}(t)=\frac{1}{2} [ {\bf L}^{\top}-{\bf L}(t) ]$.
We do not dwell, however, on such a relation, because this
work focuses on uniaxial tension of an incompressible
medium, when the tensors, ${\bf \Omega}$ and ${\bf \Omega}_{s}$,
vanish.

\section{Material functions}

The dimensionless strain energy density, $w$, is given by
\begin{equation}
w=J_{1}-3.
\end{equation}
Equation (46) describes the mechanical energy of a neo-Hookean 
medium.
The advantages of this equation are that it (i) contains no adjustable 
parameters, and (ii) has a transparent physical meaning
as the strain energy density of a Gaussian network of flexible chains
\cite{Klo02}.

The following phenomenological relation is proposed for the function
$\alpha(t)$:
\begin{equation}
\alpha =a \Bigl [ 1-\exp
\Bigl (-\frac{\sqrt{J_{1}-3}}{\varepsilon}\Bigr ) \Bigr ].
\end{equation}
Formula (47) is determined by two adjustable parameters, 
$a$ and $\varepsilon$:
the coefficient $a$ is the rate of sliding of junctions for a developed 
viscoplastic flow, 
and the strain $\varepsilon$ characterizes transition to the steady flow.
Equation (47) was successfully employed in our previous 
study on the viscoplastic response of iPP at small 
strains \cite{DC02b}.
Similar relation was suggested in \cite{KK74} to describe 
the time-dependent response of polyethylene melts.

To describe changes in the elastic modulus, $M$, 
induced by texture formation in the post-yield region 
of deformations, we adopt the phenomenological equation
\begin{equation}
M(j_{1},j_{2})=\frac{1}{2}E \Bigl [1+\eta (j_{1}-3) \Bigr ].
\end{equation}
An advantage of Eq. (48) is that it is determined by two adjustable
parameters, $E$ and $\eta$, and the effect of viscoplastic flow
on the elastic modulus is accounted in a fashion similar to that
employed in Eqs. (46) and (47).

As this work focuses on experimental data in uniaxial tensile tests,
an explicit expression for the yield surface cannot be verified.
We accept the von Mises criterion and assume that 
in the sub-yield region of deformations,
\[
\Sigma_{\rm i}<\Sigma_{\rm y},
\]
whereas in the post-yield region,
\[
\Sigma_{\rm i}\geq \Sigma_{\rm y},
\]
where
\begin{equation}
\Sigma_{\rm i}=\Bigl (\frac{3}{2}{\bf \Sigma}^{\prime}:{\bf \Sigma}^{\prime}\Bigr )^{\frac{1}{2}}
\end{equation}
is the true stress intensity.

Substituting expressions (46) and (48) into Eqs. (43) to (45)
and using Eq. (36), we find that in the sub-yield region of deformations,
\begin{eqnarray}
{\bf \Sigma}(t) &=& -p(t){\bf I}+E\Bigl [ {\bf B}_{\rm e}(t)
-\alpha(t){\bf C}_{\rm e}(t) \Bigr ],
\nonumber\\
{\bf D}_{\rm s}(t) &=& \alpha(t) {\bf D}(t),
\end{eqnarray}
and in the post-yield region of deformations,
\begin{eqnarray}
{\bf \Sigma}(t) &=& -p(t){\bf I}+E\biggl \{ \Bigl [ 1+\eta \Bigl (j_{1}(t)-3 \Bigr )\Bigr ]
\Bigl [ {\bf B}_{\rm e}(t)-\alpha(t){\bf C}_{\rm e}(t) \Bigr ]
\nonumber\\
&& +\eta \alpha(t) \Bigl (J_{1}(t)-3\Bigr ) {\bf b}(t) \biggr \},
\nonumber\\
{\bf D}_{\rm s}(t) &=& \alpha(t) {\bf D}(t)+\gamma E D_{\rm i}(t) 
\biggl \{ \Bigl [ 1+\eta \Bigl ( j_{1}(t) -3\Bigr )\Bigr ] {\bf C}_{\rm e}(t)
\nonumber\\
&& -\eta \Bigl ( J_{1}(t)-3 \Bigr ) {\bf b}(t) \biggr \}^{\prime}.
\end{eqnarray}
The stress--strain relations (50) and (51) are determined by 5 
adjustable parameters: $a$, $E$, $\gamma$, $\eta$ and $\varepsilon$.
An advantage of this model is that the number 
of constants to be found by fitting experimental data is 
substantially smaller than that in other constitutive relations 
in finite viscoplasticity of solid polymers 
\cite{BJJ96,MDW97,BSL00,HB95,BK95,BSG97,SK98,FB01}.

Our purpose now is to simplify the stress--strain relations for uniaxial
tension of an incompressible medium.

\section{Uniaxial tension}

Points of a bar refer to Cartesian coordinates $\{ X_{i} \}$  $(i=1,2,3)$
in the stress-free state,
to Cartesian coordinates $\{ x_{i} \}$ in the deformed state, 
and to Cartesian coordinates $\{ \xi_{i} \}$ in the reference state 
at time $t$. 
Uniaxial tension of the incompressible medium 
is described by the formulas
\begin{equation}
x_{1}=k(t)X_{1}, \qquad 
x_{2}=k^{-\frac{1}{2}}(t) X_{2}, \qquad
x_{3}=k^{-\frac{1}{2}}(t) X_{3},
\end{equation}
where $k=k(t)$ is an elongation ratio. 
It is assumed that transformation of the reference state is determined by
the equations similar to Eq. (52),
\begin{equation}
\xi_{1}=\kappa(t)X_{1}, \qquad 
\xi_{2}=\kappa^{-\frac{1}{2}}(t) X_{2}, \qquad
\xi_{3}=\kappa^{-\frac{1}{2}}(t) X_{3},
\end{equation}
where $\kappa(t)$ is a function to be found.
It follows from Eqs. (2), (3), (8), (52) and (53) that
\begin{equation}
{\bf B}_{\rm e}={\bf C}_{\rm e}=\Bigl (\frac{k}{\kappa}\Bigr )^{2}
{\bf e}_{1}{\bf e}_{1}
+\frac{\kappa}{k} ({\bf e}_{2}{\bf e}_{2}+{\bf e}_{3}{\bf e}_{3}),
\end{equation}
where ${\bf e}_{i}$ are base vectors of the Cartesian frame 
$\{ X_{i}\}$.
By analogy with Eq. (54), we find that in the post-yield region
of deformations,
\begin{equation}
{\bf b}={\bf c}=\Bigl ( \frac{\kappa}{\kappa_{\rm y}}\Bigr )^{2}
{\bf e}_{1}{\bf e}_{1}
+\frac{\kappa_{\rm y}}{\kappa}\Bigl ({\bf e}_{2}{\bf e}_{2}
+{\bf e}_{3}{\bf e}_{3} \Bigr ),
\end{equation}
where $\kappa_{\rm y}$ is the value of $\kappa$ at the yield point.
It follows from Eqs. (22), (54) and (55) that 
\begin{equation}
J_{1}(k,\kappa)=\Bigl ( \frac{k}{\kappa}\Bigr )^{2}+2\frac{\kappa}{k},
\qquad
j_{1}(\kappa,\kappa_{\rm y})=\Bigl ( \frac{\kappa}{\kappa_{\rm y}}\Bigr )^{2}
+2\frac{\kappa_{\rm y}}{\kappa}.
\end{equation}
Substitution of Eqs. (54) and (55) into Eqs. (50) and (51) implies that
\[
{\bf \Sigma}=\Sigma_{1}{\bf e}_{1}{\bf e}_{1}
+\Sigma_{2} ({\bf e}_{2}{\bf e}_{2}+{\bf e}_{3}{\bf e}_{3} ).
\]
In the sub-yield region of deformations,
the non-zero components of the Cauchy stress tensor, $\Sigma_{1}$
and $\Sigma_{2}$, are given by
\begin{eqnarray}
\Sigma_{1} &=& -p(k,\kappa)+E\Bigl [ 1-\alpha(k,\kappa)\Bigr ] \Bigl (\frac{k}{\kappa}\Bigr )^{2},
\nonumber\\
\Sigma_{2} &=& -p(k,\kappa)+E\Bigl [1-\alpha(k,\kappa)\Bigr ] \frac{\kappa}{k}.
\end{eqnarray}
In the post-yield region of deformations, these quantities read
\begin{eqnarray}
\Sigma_{1} &=& -p(k,\kappa,\kappa_{\rm y})+E\biggl \{ \Bigl [ 1-\alpha(k,\kappa)\Bigr ] 
\Bigl [1+\eta \Bigl (j_{1}(\kappa,\kappa_{\rm y})-3\Bigr )\Bigr ]
\Bigl (\frac{k}{\kappa}\Bigr )^{2}
\nonumber\\
&& +\eta \alpha(k,\kappa)\Bigl (J_{1}(k,\kappa)-3 \Bigr )
\Bigl (\frac{\kappa}{\kappa_{\rm y}}\Bigr )^{2}\biggr \},
\nonumber\\
\Sigma_{2} &=& -p(k,\kappa,\kappa_{\rm y})+E\biggl \{ \Bigl [ 1-\alpha(k,\kappa)\Bigr ] 
\Bigl [1+\eta \Bigl (j_{1}(\kappa,\kappa_{\rm y})-3\Bigr )\Bigr ]
\frac{\kappa}{k}
\nonumber\\
&& +\eta \alpha(k,\kappa) \Bigl (J_{1}(k,\kappa)-3 \Bigr )
\frac{\kappa_{\rm y}}{\kappa}\biggr \},
\end{eqnarray}
where the function $\alpha(k,\kappa)$ is given by Eqs. (47) and (56).

Excluding the unknown pressure, $p$, from Eqs. (57) and (58) and
the boundary condition, $\Sigma_{2}=0$, on the lateral surface of the bar, 
we find the only component of the Cauchy stress tensor, 
the true longitudinal stress $\Sigma=\Sigma_{1}$.
Bearing in mind that for uniaxial tension of an incompressible medium,
$\Sigma_{\rm i}=\Sigma$, we obtain
\begin{eqnarray}
\Sigma &=& E\Bigl [1-\alpha(k,\kappa)\Bigr ]
\biggl [ \Bigl (\frac{k}{\kappa}\Bigr )^{2}-\frac{\kappa}{k} \biggr ]
\qquad
(\Sigma<\Sigma_{\rm y}),
\nonumber\\
\Sigma &=& E\biggl \{ \Bigl [ 1-\alpha(k,\kappa)\Bigr ] 
\Bigl [1+\eta \Bigl (j_{1}(\kappa,\kappa_{\rm y})-3\Bigr )\Bigr ]
\Bigl [ \Bigl (\frac{k}{\kappa}\Bigr )^{2}-\frac{\kappa}{k} \Bigr ]
\nonumber\\
&& +\eta \alpha(k,\kappa)\Bigl (J_{1}(k,\kappa)-3 \Bigr )
\Bigl [ \Bigl (\frac{\kappa}{\kappa_{\rm y}}\Bigr )^{2}
-\frac{\kappa_{\rm y}}{\kappa} \Bigr ] \biggr \}
\qquad
(\Sigma\geq \Sigma_{\rm y}).
\end{eqnarray}
Equations (10), (13), (52) and (53) result in the formulas
\begin{equation}
{\bf D}=\frac{1}{k}\frac{dk}{dt} \Bigl [ {\bf e}_{1}{\bf e}_{1}
-\frac{1}{2} ({\bf e}_{2}{\bf e}_{2}
+{\bf e}_{3}{\bf e}_{3} ) \Bigr ],
\qquad
{\bf D}_{\rm s}=\frac{1}{\kappa}\frac{d\kappa}{dt} 
\Bigl [ {\bf e}_{1}{\bf e}_{1}
-\frac{1}{2} ({\bf e}_{2}{\bf e}_{2}
+{\bf e}_{3}{\bf e}_{3} ) \Bigr ] ,
\end{equation}
which, together with Eq. (42), imply that
\begin{equation}
D_{\rm i}= \frac{1}{k}\frac{dk}{dt}.
\end{equation}
It follows from Eqs. (50), (51), (60) and (61) that
\begin{eqnarray}
\frac{d\kappa}{dk} &=& \alpha(k,\kappa) \frac{\kappa}{k}
\qquad
(\Sigma<\Sigma_{\rm y}),
\nonumber\\
\frac{d\kappa}{dk} &=& \frac{\kappa}{k} \biggl \{ \alpha(k,\kappa)
+\Gamma 
\biggl [ \Bigl (1+\eta \Bigl (j_{1}(\kappa,\kappa_{\rm y})-3\Bigr )\Bigr )
\Bigl ( \Bigl (\frac{k}{\kappa}\Bigr )^{2}-\frac{\kappa}{k} \Bigr )
\nonumber\\
&& -\eta \Bigl (J_{1}(k,\kappa)-3\Bigr )
\Bigr (\Bigr (\frac{\kappa}{\kappa_{\rm y}}\Bigr )^{2}
-\frac{\kappa_{\rm y}}{\kappa}\Bigr )\biggr ]\biggr \}
\qquad (\Sigma\geq \Sigma_{\rm y})
\end{eqnarray}
with $\Gamma=\frac{2}{3}\gamma E$.

Given a loading program, $k=k(t)$, the longitudinal stress, $\Sigma(t)$,
is determined by Eq. (59).
The elongation ratio, $\kappa(t)$, that characterizes fine and coarse 
slips of lamellar blocks, is found from the nonlinear differential
equations (62) with the initial condition $\kappa(1)=1$.
The constitutive equations involve 5 adjustable parameters:
\begin{enumerate}
\item
the elastic modulus $E$,

\item
the rate of a developed viscoplastic flow of junctions $a$,

\item
the strain $\varepsilon$ that characterizes transition 
to a steady flow of junctions,

\item
the rate of coarse slip of lamellar blocks $\Gamma$,

\item
the coefficient $\eta$ that characterizes strain-hardenig
in the post-yield region of deformations.
\end{enumerate}
The yield stress, $\Sigma_{\rm y}$, is determined directly 
from a stress--strain diagram as the true stress corresponding to
the point of maximum on the engineering stress--engineering strain
curve.
To find the viscoplastic elongation ratio at yielding, $\kappa_{\rm y}$, 
the constitutive equations are integrated from $\Sigma=0$ 
to $\Sigma=\Sigma_{\rm y}$.

An important advantage of Eqs. (59) and (62) is that
3 material constants, $E$, $a$ and $\varepsilon$, are found
by fitting a stress--strain curve below the yield point.
Afterwards, the other two parameters, $\Gamma$ and $\eta$, are 
determined by matching the stress--strain curve 
in the post-yield region of deformations.

\section{Fitting of observations}

We begin with matching the stress--strain diagrams 
below the apparent yield point.
To find the constants $E$, $a$ and $\varepsilon$,
we fix some intervals $[0,a_{\max}]$ 
and $[0,\varepsilon_{\max}]$, where the ``best-fit" parameters 
$a$ and $\varepsilon$ are assumed to be located,
and divide these intervals into $J$ subintervals by
the points $a^{(i)}=i\Delta a$ and $\varepsilon^{(j)}=j\Delta \varepsilon$  
($i,j=1,\ldots,J$) with $\Delta a=a_{\max}/J$ and 
$\Delta \varepsilon=\varepsilon_{\max}/J$.
For any pair, $\{ a^{(i)}, \varepsilon^{(j)} \}$, 
Eqs. (59) and (62) are integrated numerically by the Runge--Kutta method
with the step $\Delta k=1.0\cdot 10^{-5}$ in the
interval between $\Sigma=0$ and $\Sigma=\Sigma_{\rm y}$.
Given a pair, $\{ a^{(i)}, \varepsilon^{(j)} \}$, the elastic modulus, 
$E$, is found by the least-squares method
from the condition of minimum of the function
\[
R=\sum_{k_{n}} \Bigl [ \Sigma_{\rm exp}(k_{n})
-\Sigma_{\rm num}(k_{n}) \Bigr ]^{2},
\]
where the sum is calculated over all experimental points, $k_{n}$, 
in the sub-yield interval,
$\Sigma_{\rm exp}$ is the stress measured in a tensile test, 
and $\Sigma_{\rm num}$ is given by Eq. (59).
The ``best-fit" parameters $a$ and $\varepsilon$ are determined
from the condition of minimum of the function $R$ 
on the set $ \{ a^{(i)}, \varepsilon^{(j)} \quad (i,j=1,\ldots, J)  \}$.

The material constants $E$, $a$ and $\varepsilon$ that
minimize the discrepancies between the experimental data
and the results of numerical analysis are found for any
stress--strain curve independently.

First, we approximate the observations for non-annealed 
specimens.
Figures 1 to 7 demonstrate good quality of fitting 
the stress--strain curves up to the points of necking formation.
This result is rather surprising.
In terms of the model, it means that for samples
not subjected to thermal treatment, necking (transition from
a homogeneous to a spatially heterogeneous deformation
of samples) precedes yielding (coarse slip and fragmentation
of lamellar blocks).
This implies that the presence of a wide plateau on a stress--strain
diagram for a non-annealed specimen near the point of maximum
may be attributed not to the material yielding (as conventional
scenarios for yielding suggest), but to a developed flow of junctions
in the amorphous phase and fine slip of lamellar blocks 
in the crystalline phase without lamellar fragmentation and texture 
formation.

Afterwards, we match the stress--strain curves in the sub-yield
region of deformations for annealed specimens.
Figures 8 to 14 show fair agreement between the experimental
data and the results of numerical analysis.

The adjustable parameters $E$, $a$ and $\varepsilon$ are plotted
versus the rate of Hencky strain, $\dot{\epsilon}_{\rm H}$, in
Figures 16 to 18.
The experimental data are fitted by the functions
\begin{equation}
E=E_{0}+E_{1}\log \dot{\epsilon}_{\rm H},
\qquad
a=a_{0}+a_{1}\log \dot{\epsilon}_{\rm H},
\qquad
\log \varepsilon=\varepsilon_{0}
+\varepsilon_{1}\log \dot{\epsilon}_{\rm H},
\end{equation}
where the coefficients $E_{m}$, $a_{m}$ and 
$\varepsilon_{m}$ ($m=0,1$) are found by the least-square 
technique.
Figures 16 to 18 demonstrate acceptable agreement between
the observations and their approximations by phenomenological
relations (63).
It is worth noting rather large scatter of the experimental data for
annnealed samples compared to that for specimens not subjected
to thermal pre-treatment.
These discrepancies may be explained by a substantial decrease in 
the interval of elongation ratios where the fitting procedure
is performed (the maximal engineering strain for the interval where 
the stress--strain curves are approximated is reduced by a
factor of three: from 0.15 to 0.05).

To find the quantities $\Gamma$ and $\eta$, we approximate 
the stress--strain curves for annealed specimens
in the post-yield region of deformation.
For any set of experimental data, we use the parameters
$E$, $a$ and $\varepsilon$ found by fitting an 
appropriate stress--strain curve in the sub-yield region
of deformations.
To approximate a stress--strain curve above the yield point,
we apply an algorithm similar to that used to match 
the observations in the sub-yield region of deformations.
We fix some intervals $[0,\Gamma_{\max}]$ and $[0,\eta_{\max}]$,
where the ``best-fit" parameters $\Gamma$ and $\eta$
are assumed to be located,
and divide these intervals into $J$ subintervals by
the points $\Gamma^{(i)}=i\Delta \Gamma$
and $\eta^{(j)}=j\Delta \eta$ ($i,j=1,\ldots,J$)
with $\Delta \Gamma=\Gamma_{\max}/J$
and $\Delta \eta=\eta_{\max}/J$.
Given a pair, $\{ \Gamma^{(i)}, \eta^{(j)} \}$, 
Eqs. (59) and (62) are integrated numerically 
by the Runge--Kutta method with the step $\Delta k=1.0\cdot 10^{-5}$.
The ``best-fit" parameters $\Gamma$ and $\eta$ 
are found from the condition of minimum of the function $R$ 
on the set $ \{ \Gamma^{(i)}, \eta^{(j)} \quad (i,j=1,\ldots, J)  \}$.

The dimensionless parameters, $\Gamma$ and $\eta$, are plotted
versus the Hencky strain rate, $\dot{\epsilon}_{\rm H}$,
in Figures 19 and 20.
The observations for the rate of coarse slip of lamellar blocks
show that $\Gamma$ is constant, $\Gamma=\Gamma_{0}$.
The experimental data for the parameter $\eta$
are approximated by the function
\begin{equation}
\eta=\eta_{0}+\eta_{1}\log \dot{\epsilon}_{\rm H},
\end{equation}
where the coefficients $\eta_{m}$ ($m=0,1$) 
are found by the least-squares algorithm.
Figure 20 demonstrates that Eq. (64) ensures acceptable quality 
of matching the observations.

\section{Discussion}

Figures 1 to 14 demonstrate fair agreement between the 
experimental data in uniaxial tensile tests and the results 
of numerical simulation, which means that the constitutive 
equations can be successfully applied to fit observations 
for isotactic polypropylene (non-annealed as well as subjected 
to thermal pre-treatment) at various strain rates and elongation ratios.
This conclusion contradicts the conventional standpoint \cite{Klo02} 
that (i) the concept of Gaussian networks of flexible chains 
(which implies Eq. (46) for the mechanical energy per unit volume) 
fails to correctly approximate stress--strain curves for
solid polymers at finite strains 
and (ii) more sophisticated expressions are necessary for 
the strain energy density $w$ 
(based on either the slip-link theory \cite{SCC97,SCC99,MDW97}
or a multi-chain version of the theory of polymeric networks
with finite extensibility of chains \cite{BKR02,LB99,BSL00,HB95}).
Within the model proposed, an acceptable quality of matching
observations is reached based on the hypotheses about
(i) non-affine motion of nodes in the network 
and (ii) the growth of the elastic modulus driven by 
texture formation.

Figure 16 reveals that the elastic modulus, $E$,
monotonically increases with the rate of stretching for 
non-annealed specimens and decreases for annealed samples.
The following explanations may be provided for this finding.

As the present study concentrates on the viscoplastic behavior
of semicrystalline polymers, we disregard the viscoelastic phenomena 
associated with rearrangement of chains 
(separation of active strands from junctions
and merging of dangling strands with the network)
and treat a polymer as an equivalent network of 
chains bridged by permanent junctions.
Given a strain rate, the simplest way to account for these effects 
is to distinguish between two groups of meso-regions in an ensemble:
\begin{itemize}
\item
the characteristic time for rearrangement of chains in MRs
belonging to the first group is substentially smaller than the 
characteristic time of loading (which means that the contribution 
of these meso-domains into the strain energy of a polymer is
negligible), 

\item
the characteristic time for rearrangement of chains in MRs 
of the other group exceeds the characteristic time for loading
(which implies that rearrangement of chains in these meso-domains
may be disregarded and they may be considered as permanent
networks).
\end{itemize}
According to this division of MRs into two groups,
the number of meso-regions with permanent junctions per 
unit volume, $N$, in the formula for the elastic modulus, $E$, 
grows with the Hencky strain rate $\dot{\epsilon}_{\rm H}$
(because an increase in $\dot{\epsilon}_{\rm H}$
results in a decrease in the characteristic time of loading,
and, as a consequence, a decrease in the content of
MRs where stresses in chains relax due to the rearrangement
process).
This conclusion is in agrement with the experimental data 
presented in Figure 16 for non-annealed specimens (curve 1).
When the cross-head speed decreases from 200 to 5 mm/min 
(which corresponds to an increase in the loading time from 4 to 180 s), 
the elastic modulus, $E$, is reduced by about 15\%, 
which is quite comparable with the intensity of decay in
the longitudinal stress during 180 s in uniaxial tensile 
relaxation tests \cite{DC02a}.

The substantial decrease in the elastic modulus of annealed
specimens with the Hencky strain rate (curve 2 in Figure 16) may be
associated with alteration of the morphological structure of iPP
at thermal treatment.

Stretching of a semicrystalline polymer in the sub-yield region 
of deformations activates sliding of junctions in amorphous 
domains between lamellae.
Displacements of junctions induce separation of tie 
chains from crystallites and their slippage along lamellar 
blocks.
When the tie chains are rather long, they have enough time
to attach to lamellae in new cites, which implies that the
total number of links between amorphous regions and
crystalline lamellae (than transmit the macro-strain to
amorphous domains) remains practically constant.
The latter means that amorphous meso-domains do not
separate from the ensemble at stretching of a non-annealed
specimen.

The situation changes dramatically when a semicrystalline polymer
is subjected to thermal pre-treatment.
Annealing of iPP for 51 h at 160 $^{\circ}$C results in a
noticeable increase in the degree of crystallinity.
Crystallization of polymeric chains in amorphous regions 
occurs strongly inhomogeneously: long tie strands 
(that have high molecular mobility) are mainly crystallized, 
whereas majority of short tie chains remain in the amorphous state.
This implies that the average length of a tie strand substantially 
decreases at thermal treatment.
Sliding of junctions in amorphous domains caused by straining
of a specimen induces separation of tie chains from crystallites 
and their slippage along lamellar surfaces.
Unlike non-annealed specimens, where tie chains are rather long
(which implies that their mobility is high and the time necessary for
their reconnection is small), for annealed samples, 
the rate of attachment of short tie chains (with low molecular mobility) 
to new cites on the surfaces of crystallites becomes comparable 
with the strain rate.
This means that an increase in the Hencky strain rate, 
$\dot{\epsilon}_{\rm H}$, results in a strong decrease in 
the concentration of links that transmit the macro-strain 
to amorphous domains.
As a consequence, individual MRs separate from the ensemble
(due to the breakage of links between amorphous 
meso-domains and surrounding regions), which implies that 
the number of ``active" meso-regions per unit volume, $N$, 
decreases with $\dot{\epsilon}_{\rm H}$.
As the elastic modulus, $E$, is proportional to $N$, the growth
of the strain rate induces a decrease the elastic modulus of
annealed specimens, in accord with the experimental data 
depicted in Figure 16 (curve 2).

It is worth noting that at small strain rates (when detachment
of tie chains from lamellar blocks is not pronounced)
the elastic modulus, $E$, of samples subjected to thermal
treatment substantially (about by twice) exceeds that 
of non-annealed specimens (due to the growth in the degree 
of crystallinity).
With an increase in the Hencky strain rate, the difference between
the moduli of annealed and non-annealed samples is reduced,
and at relatively high strain rates (above $\dot{\epsilon}_{\rm c}\approx
5 \cdot 10^{-2}$ s$^{-1}$), the elastic modulus of non-annealed
specimens exceeds that of annealed ones.

Figure 17 shows that the dimensionless rate of a developed flow 
of junctions, $a$, monotonically increases with the strain rate.
This phenomenon may be ascribed to breakage of tie chains
that bridge amorphous regions with lamellar blocks.
A decrease in the concentration of these chains implies that the
mobility of junctions in amorphous MRs grows, 
which is observed as an increase in $a$ with $\dot{\epsilon}_{\rm H}$.
At small strain rates (below the threshold value, $\dot{\epsilon}_{\rm c}$),
the rate of a steady viscoplastic flow, $a$, for non-annealed
specimens exceeds that for samples subjected to thermal
treatment (this observation is ascribed to the secondary 
crystallization at annealing that reduces molecular mobility
in meso-regions),
whereas at high strain rates (that exceed $\dot{\epsilon}_{\rm c}$),
the inverse picture is revealed (the rate of a developed flow of 
junctions for annealed specimens reaches its ultimate value,
$a=1$).
In agreement with our explanation for the decrease in the
elastic modulus, $E$, of annealed specimens 
with $\dot{\epsilon}_{\rm H}$, this
result is attributed to breakage of tie chains (which have no time 
enough for their reconnection).
A decrease in the content of tie chains implies a strong increase in
molecular mobility in amorphous meso-regions, which is reflected
by the model as the growth of the rate of viscoplastic flow $a$.

According to Figure 18, the strain, $\varepsilon$, that characterizes
transition to a developed flow of junctions, increases with the
rate of straining both for non-annealed and annealed specimens.
It is worth noting that the increase in $\varepsilon$ is more
pronounced for annealed samples than for non-annealed ones 
(however, in the former case, the scatter of experimental data is
rather large).

Figure 19 demonstrates that the rate, $\Gamma$, of coarse slip 
and fragmentation of lamellar blocks in the post-yield region of
deformations is independent of the strain rate.
This results appears to be quite natural: it means that the energy necessary
for disintegration of lamellar blocks exceeds substentially the
elastic energy stored in the semicrystalline polymer,
which implies that the rate of coarse slip is not affected
noticeably by mechanical factors.
These factors do affect, however, the process of lamellar fragmentation,
because, in accord with Eqs. (59) and (62), the derivative, $d\kappa/dk$, is
proportional to the longitudinal stress, $\Sigma$.

Figure 20 reveals that the dimensionless parameter $\eta$ (that 
characterizes strain-hardening of a semicrystalline polymer) 
monotonocally increases
with the Hencky strain rate, $\dot{\epsilon}_{\rm H}$.
This assertion appears to be quite natural.
Figure 16 shows that the content of tie chains in annealed 
specimens decreases with the strain rate.
As a consequence of this reduction in the concentration
of tie chains, broken lamellar blocks 
become more mobile, and they are oriented easier 
along the direction of the longitudinal stress.
The strain-rate induced intensification of the alignment 
process for disintegrated lamellar blocks implies 
a more pronounced increase in the stress, $\Sigma$, 
with the elongation ratio, $k$, in the post-yield
region of deformations observed in Figure 20.

According to Eq. (48) and Figure 20, the growth of 
the elastic modulus, $E$, driven by strain-hardening of 
iPP (in tensile tests with the maximal Hencky strain 
$\epsilon_{\rm H}=0.6$) is of order of few per cent.
It is worth noting that the same (qualitatively) level of
orientation of lamellar blocks was recently observed
in SAXS (small angle X-ray scattering) measurements
on another semicrystalline polymer (polyamide PA11) 
at uniaxial tension up to the necking point \cite{JTH02}.

\section{Concluding remarks}

Two series of uniaxial tensile tests have been performed 
on isotactic polypropylene at room temperature.
The first series of experiments is carried out on 
injection-molded specimens non-subjected to thermal treatment.
In the other series of tests, the samples were annealed
for 51 h at the temperature $T=160$~$^{\circ}$C prior to loading.
Each series consists of 7 experiments with the cross-hear speeds 
ranging from 5 to 200 mm/min
(which cover practically the entire range of cross-head speeds
employed in conventional quasi-static tensile tests).
Stretching of a non-annealed specimen is performed up to
the onset of necking.
Deformation of a sample subjected to thermal treatment is
carried out up to the maximal Hencky strain $\epsilon_{\rm H}=0.6$.

Constitutive equations have been derived for the isothermal 
viscoplastic behavior of a semicrystalline polymer at finite strains.
A polymer is treated as an equivalent heterogeneous network 
of chains bridged by permanent junctions.
The network is thought of as an ensemble of meso-regions linked 
with each other.
Under loading, junctions between chains in MRs move with 
respect to their reference positions (this process reflects sliding 
of nodes in the amorphous phase and fine slip of lamellar blocks).
At relatively large elongations, the sliding process is accompanied 
by displacement of meso-regions with respect to each other
(which reflects coarse slip and fragmentation of lamellar blocks).

Unlike the conventional definition of the yield point of a solid
polymer at uniaxial tension as the point of maximum on
the engineering stress--engineering strain diagram,
we define the yield strain as the strain at which coarse slip
and fragmentation of lamellar blocks starts.
This implies that sliding of junctions in MRs with respect to their
initial positions occurs at any elongation ratio,
whereas sliding of meso-domains takes place in the post-yield 
region of deformations only.

Stress--strain relations and kinetic equations for coarse
slip of lamellar blocks are developed by using 
the laws of thermodynamics.
These equations are simplified for uniaxial tension
of an incompressible medium.
The governing equations are determined by 5 adjustable 
parameters that are found by fitting observations.
Fair agreement is demonstrated between the experimental
data and the results of numerical simulation both for 
non-annealed and annealed specimens.

The main result of the study is that the stress--strain
curves (up to the necking points) for specimens 
not subjected to thermal treatment can be fairly well 
approximated by the constitutive relations where 
coarse slip and fragmentation of lamellar blocks are not 
taken into account.
In terms of the model, this assertion means that necking 
of non-annealed samples occurs in the sub-yield region 
of deformations and precedes their yielding.
On the contrary, annealed specimens 
demonstrate pronounced yield points at relatively small
elongation ratios, far below the strains corresponding
to the onset of necking.

The following conclusions have been drawn:
\begin{enumerate}
\item
The elastic modulus, $E$, monotonically grows with the Hencky
strain rate, $\dot{\epsilon}_{\rm H}$, for non-annealed
specimens and monotonically decreases for annealed ones.
The difference in the mechanical response of these 
samples is attributed to a dramatic decrease 
in the average length of tie strands at thermal treatment.

\item
The rate of a developed viscoplastic flow $a$ (that reflects 
sliding of junctions in amorphous regions and fine slip of
crystalline blocks) and the strain $\varepsilon$ (that characterizes 
transition to the steady flow of junctions) increase with
the strain rate.
These observations are associated with mechanically-induced
breakage of tie chains that results in the growth 
of molecular mobility in amorphous meso-regions.

\item
The rate, $\Gamma$, of coarse slip and fragmentation of lamellar 
blocks in the post-yield region of deformations
is independent of mechanical factors.

\item
The dimensionless parameter $\eta$ 
(that characterizes strain-hardening of a semicrystalline polymer 
above the yield point) grows with the Hencky strain rate.
This observation is explained by enhancement of breakage 
of tie chains with an increase in the rate of straining,
and, as a consequence, by an acceleration of 
alignment of disintegrated lamellar blocks along the direction 
of longitudinal stress.
\end{enumerate}

\newpage

\section*{List of figures}
\parindent 0 mm

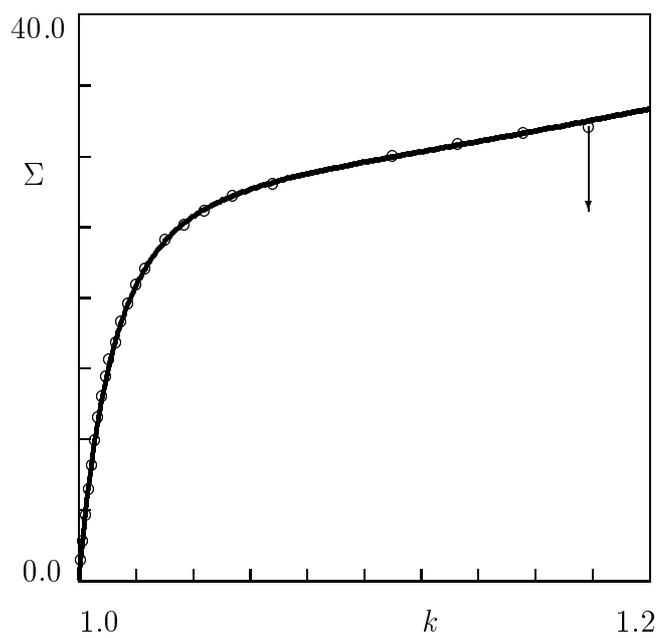
{\bf Figure 1:}
The true stress $\Sigma$ MPa versus 
the elongation ratio $k$ in a tensile test with the cross-head 
speed 5 mm/min.
Circles: experimental data for a non-annealed specimen.
Vector indicates the beginning of necking.
Solid line: results of numerical simulation
\vspace*{2 mm}

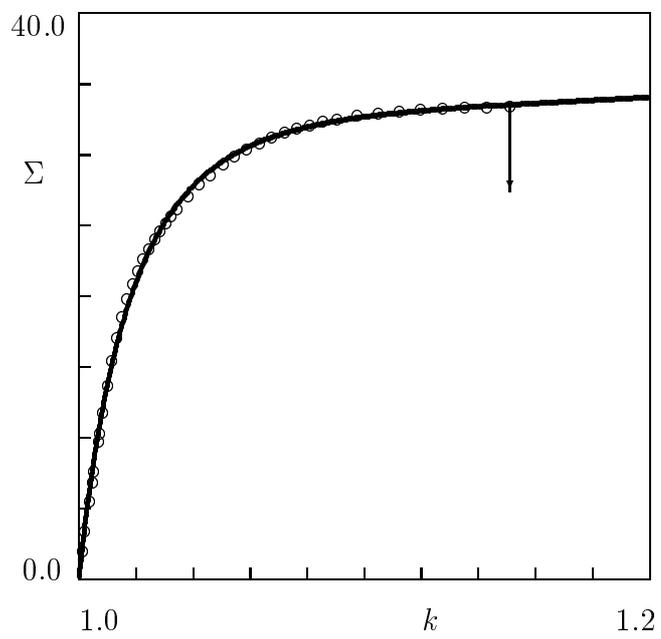
{\bf Figure 2:}
The true stress $\Sigma$ MPa versus 
the elongation ratio $k$ in a tensile test with the cross-head 
speed 10 mm/min.
Circles: experimental data for a non-annealed specimen.
Vector indicates the beginning of necking.
Solid line: results of numerical simulation
\vspace*{2 mm}

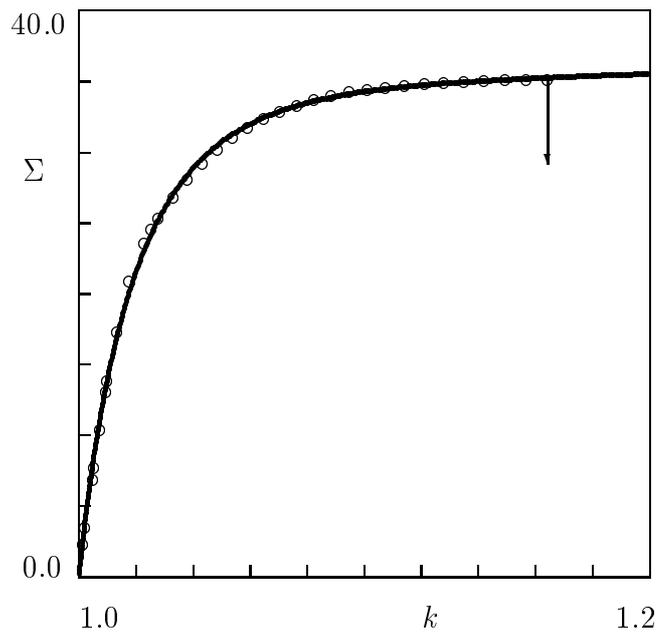
{\bf Figure 3:}
The true stress $\Sigma$ MPa versus 
the elongation ratio $k$ in a tensile test with the cross-head 
speed 25 mm/min.
Circles: experimental data for a non-annealed specimen.
Vector indicates the beginning of necking.
Solid line: results of numerical simulation
\vspace*{2 mm}

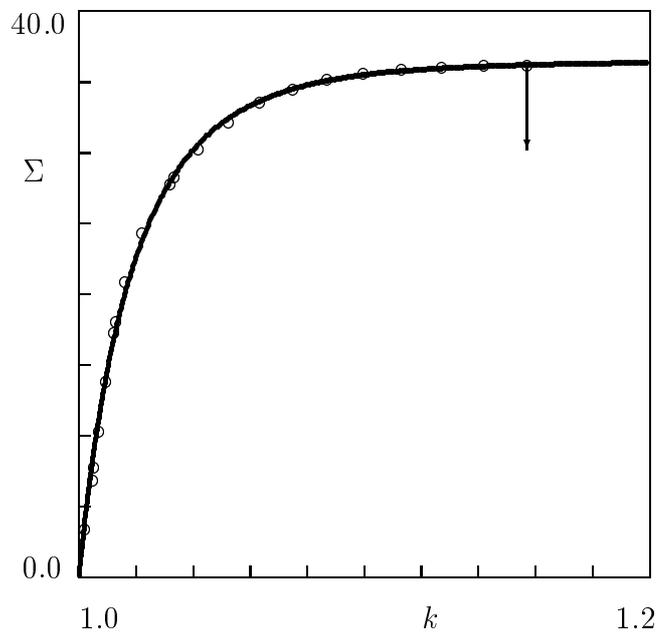
{\bf Figure 4:}
The true stress $\Sigma$ MPa versus 
the elongation ratio $k$ in a tensile test with the cross-head 
speed 50 mm/min.
Circles: experimental data for a non-annealed specimen.
Vector indicates the beginning of necking.
Solid line: results of numerical simulation
\vspace*{2 mm}

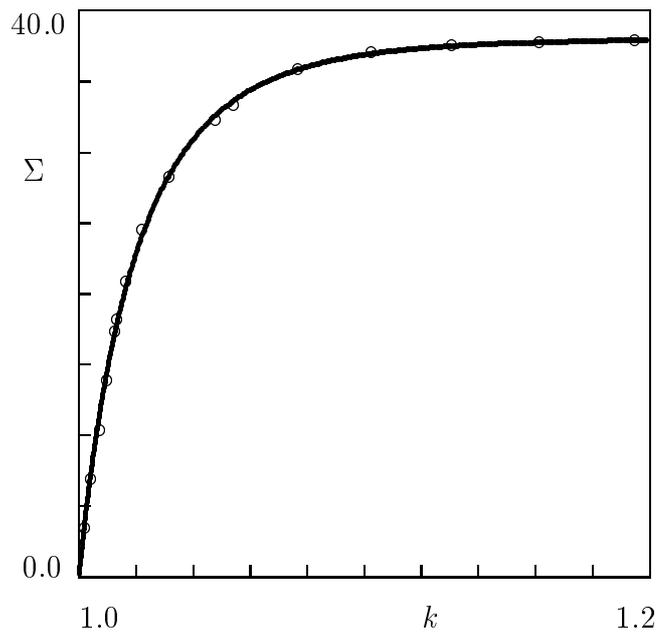
{\bf Figure 5:}
The true stress $\Sigma$ MPa versus 
the elongation ratio $k$ in a tensile test with the cross-head 
speed 100 mm/min.
Circles: experimental data for a non-annealed specimen.
Solid line: results of numerical simulation
\vspace*{2 mm}

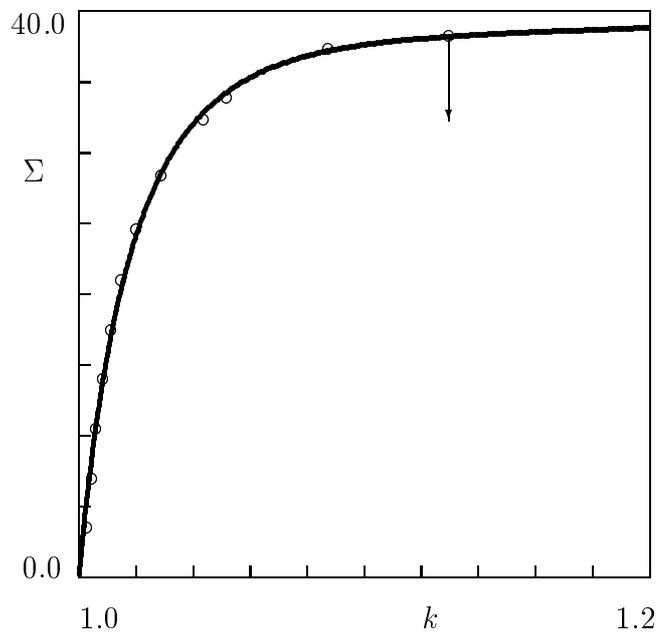
{\bf Figure 6:}
The true stress $\Sigma$ MPa versus 
the elongation ratio $k$ in a tensile test with the cross-head 
speed 150 mm/min.
Circles: experimental data for a non-annealed specimen.
Vector indicates the beginning of necking.
Solid line: results of numerical simulation
\vspace*{2 mm}

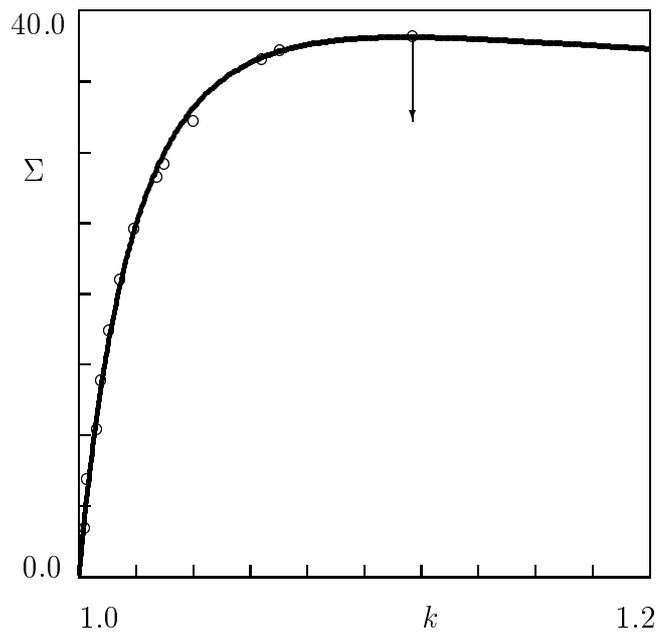
{\bf Figure 7:}
The true stress $\Sigma$ MPa versus 
the elongation ratio $k$ in a tensile test with the cross-head 
speed 200 mm/min.
Circles: experimental data for a non-annealed specimen.
Vector indicates the beginning of necking.
Solid line: results of numerical simulation
\vspace*{2 mm}

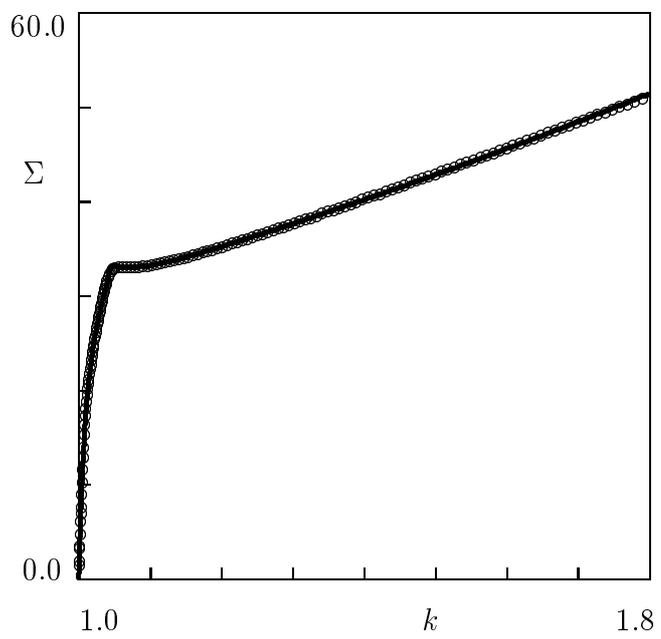
{\bf Figure 8:}
The true stress $\Sigma$ MPa versus 
the elongation ratio $k$ in a tensile test with the cross-head 
speed 5 mm/min.
Circles: experimental data for an annealed specimen.
Solid line: results of numerical simulation
\vspace*{2 mm}

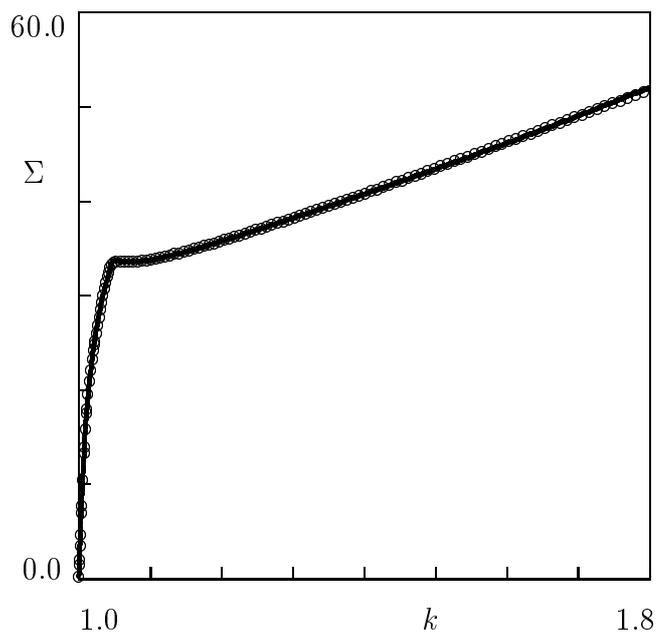
{\bf Figure 9:}
The true stress $\Sigma$ MPa versus 
the elongation ratio $k$ in a tensile test with the cross-head 
speed 10 mm/min.
Circles: experimental data for an annealed specimen.
Solid line: results of numerical simulation
\vspace*{2 mm}

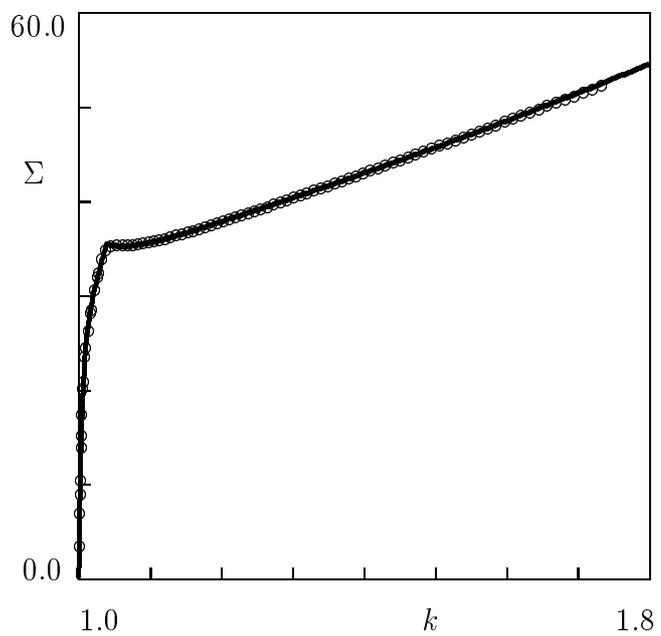
{\bf Figure 10:}
The true stress $\Sigma$ MPa versus 
the elongation ratio $k$ in a tensile 
test with the cross-head speed 25 mm/min.
Circles: experimental data for an annealed specimen.
Solid line: results of numerical simulation
\vspace*{2 mm}

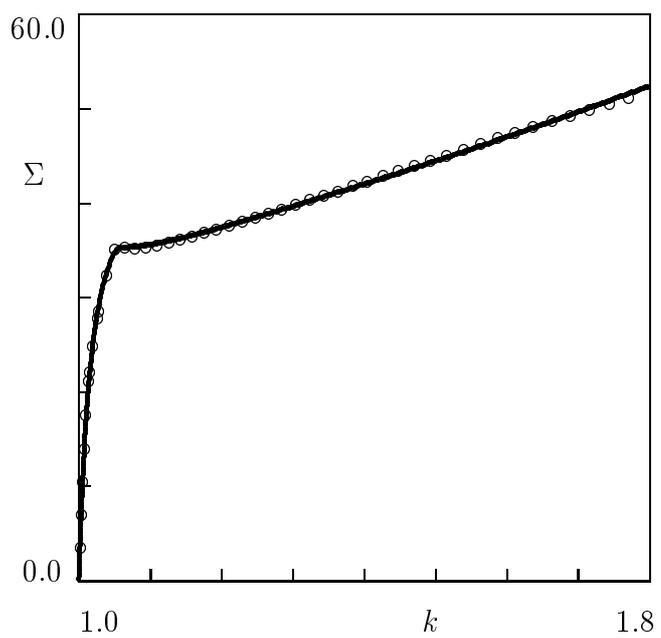
{\bf Figure 11:}
The true stress $\Sigma$ MPa versus 
the elongation ratio $k$ in a tensile 
test with the cross-head speed 50 mm/min.
Circles: experimental data for an annealed specimen.
Solid line: results of numerical simulation
\vspace*{2 mm}

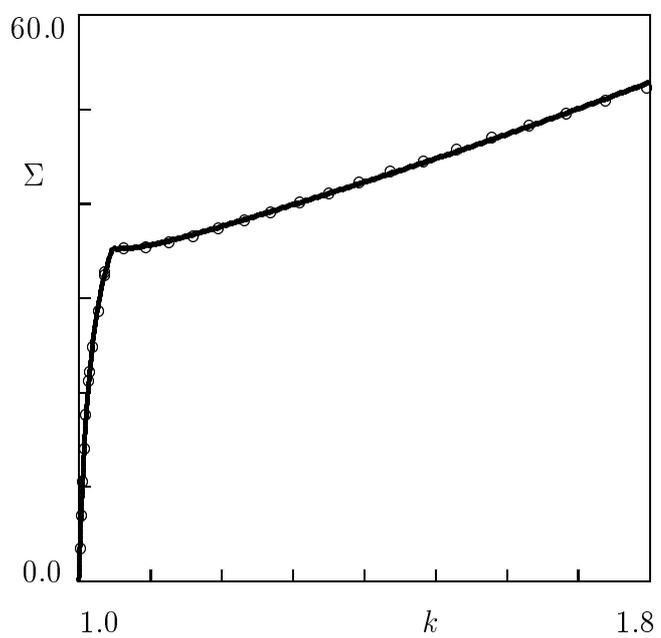
{\bf Figure 12:}
The true stress $\Sigma$ MPa versus 
the elongation ratio $k$ in a tensile 
test with the cross-head speed 100 mm/min.
Circles: experimental data for an annealed specimen.
Solid line: results of numerical simulation
\vspace*{2 mm}

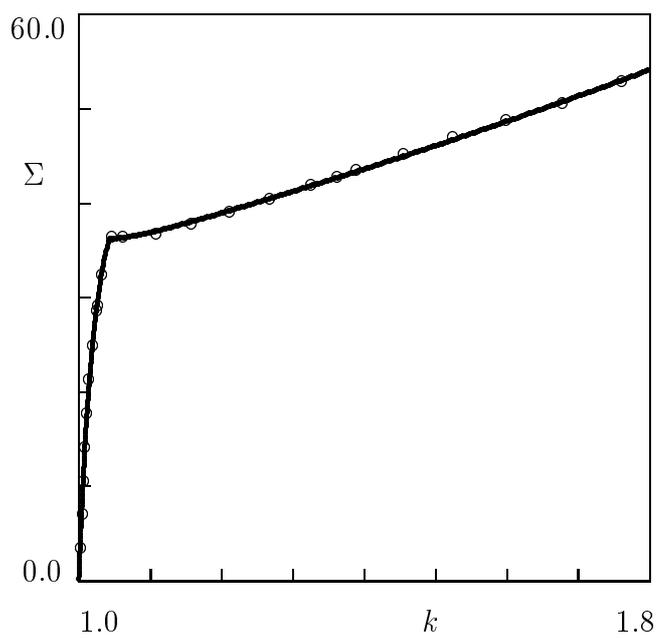
{\bf Figure 13:}
The true stress $\Sigma$ MPa versus 
the elongation ratio $k$ in a tensile 
test with the cross-head speed 150 mm/min.
Circles: experimental data for an annealed specimen.
Solid line: results of numerical simulation
\vspace*{2 mm}

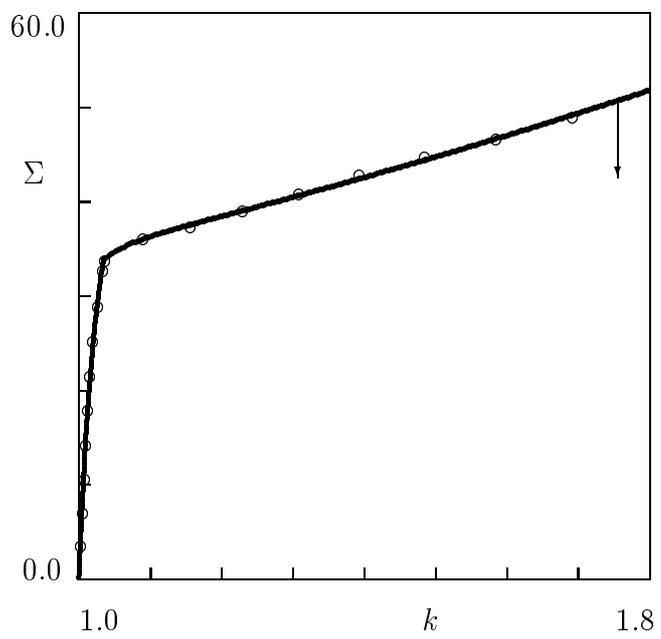
{\bf Figure 14:}
The true stress $\Sigma$ MPa versus 
the elongation ratio $k$ in a tensile 
test with the cross-head speed 200 mm/min.
Circles: experimental data for an annealed specimen.
Vector indicates the beginning of necking.
Solid line: results of numerical simulation
\vspace*{2 mm}

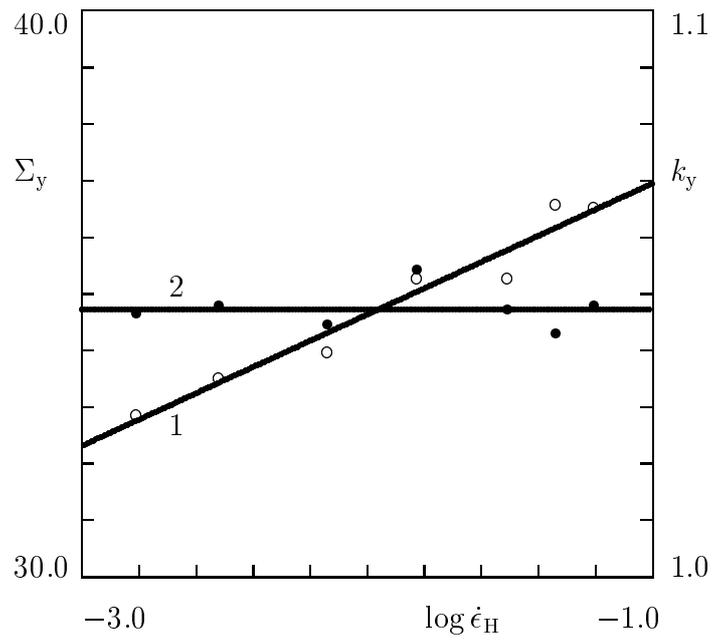
{\bf Figure 15:}
The true yield stress $\Sigma_{\rm y}$ MPa (unfilled circles)
and the elongation ratio for yielding $k_{\rm y}$ (filled circles)
versus the strain rate $\dot{\epsilon}_{\rm H}$ s$^{-1}$.
Symbols: treatment of observations.
Solid lines: approximation of the experimental data by Eq. (1)
with $\Sigma_{0}=39.27$ and $\Sigma_{1}=2.32$ (curve 1)
and by the constant $k_{\rm y}=1.0473$ (curve 2)
\vspace*{2 mm}

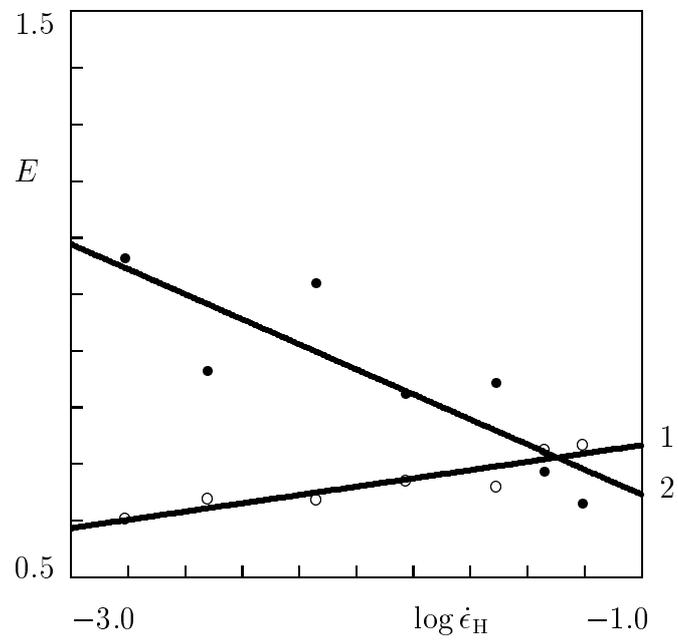
{\bf Figure 16:}
The elastic modulus $E$ GPa versus the strain rate
$\dot{\epsilon}_{\rm H}$ s$^{-1}$.
Symbols: treatment of observations.
Unfilled circles: non-annealed specimens;
filled circles: annealed specimens.
Solid lines: approximation of the experimental data by Eq. (63).
Curve 1: $E_{0}=0.81$, $E_{1}=0.07$;
curve 2: $E_{0}=0.43$, $E_{1}=-0.22$
\vspace*{2 mm}

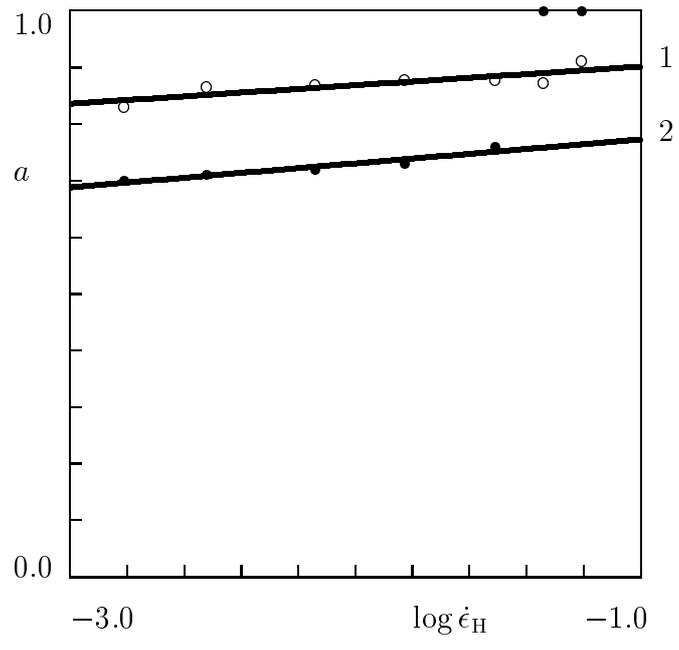
{\bf Figure 17:}
The dimensionless rate $a$ of a developed flow of 
junctions versus the strain rate $\dot{\epsilon}_{\rm H}$ s$^{-1}$.
Symbols: treatment of observations.
Unfilled circles: non-annealed specimens;
filled circles: annealed specimens.
Solid lines: approximation of the experimental data by Eq. (63).
Curve 1: $a_{0}=0.93$, $a_{1}=0.033$;
curve 2: $a_{0}=0.82$, $a_{1}=0.042$
\vspace*{2 mm}

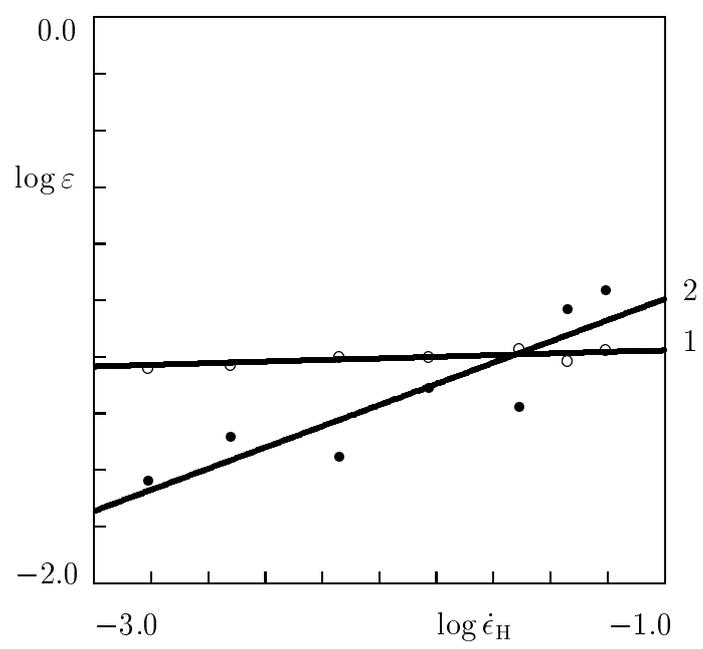
{\bf Figure 18:}
The strain $\varepsilon$ that characterizes transition
to a developed flow of junctions versus 
the strain rate $\dot{\epsilon}_{\rm H}$ s$^{-1}$.
Symbols: treatment of observations.
Unfilled circles: non-annealed specimens;
filled circles: annealed specimens.
Solid lines: approximation of the experimental data by Eq. (63).
Curve 1: $\varepsilon_{0}=-1.15$, $\varepsilon_{1}=0.028$;
curve 2: $\varepsilon_{0}=-0.62$, $\varepsilon_{1}=0.375$
\vspace*{2 mm}

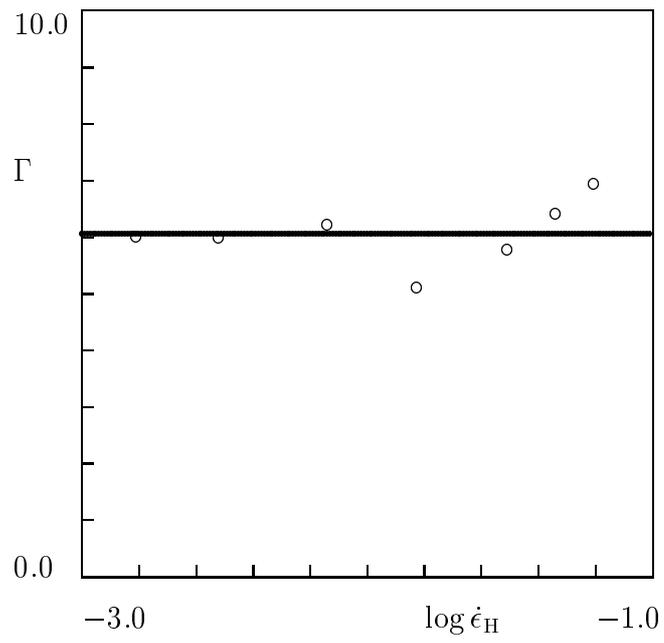
{\bf Figure 19:}
The dimensionless rate of coarse slip
of lamellar blocks $\Gamma$ versus the strain 
rate $\dot{\epsilon}_{\rm H}$ s$^{-1}$.
Circles: treatment of observations for annealed specimens.
Solid line: approximation of the experimental data by the 
constant $\Gamma_{0}=6.07$
\vspace*{2 mm}

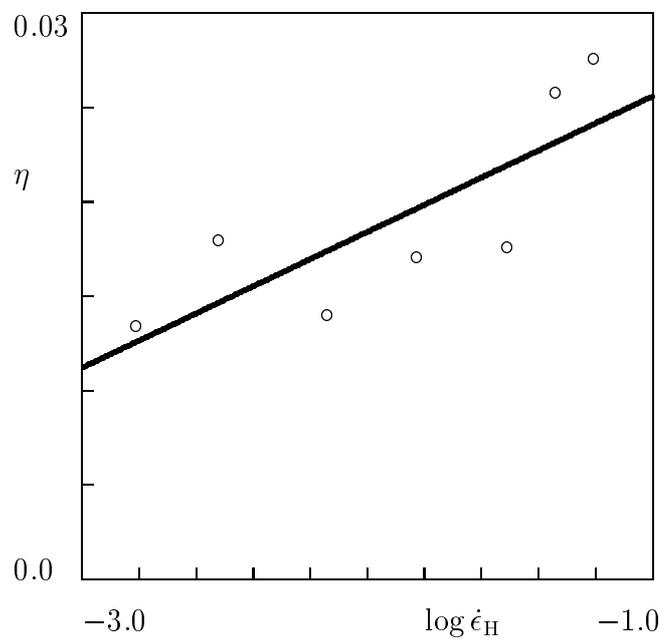
{\bf Figure 20:}
The dimensionless parameter $\eta$ that characterizes
strain-hardening versus the strain rate $\dot{\epsilon}_{\rm H}$ s$^{-1}$.
Circles: treatment of observations for annealed specimens.
Solid line: approximation of the experimental data by Eq. (64)
with $\eta_{0}=0.0328$ and $\eta_{1}=0.0072$

\newpage
\begin{thebibliography}{50}

\bibitem{BUD97}
Brooks NW, Unwin AP, Duckett RA, Ward IM.
J Polym Sci B: Polym Phys 1997; 35: 545--552.

\bibitem{GS97}
Gaucher-Miri V, Seguela R.
Macromolecules 1997; 30: 1158--1167.

\bibitem{BDW98}
Brooks NWJ, Duckett RA, Ward IM.
J Polym Sci B: Polym Phys 1998; 36: 2177--2189.

\bibitem{HHL99}
Hiss R, Hobeika S, Lynn C, Strobl G.
Macromolecules 1999; 32: 4390--4403.

\bibitem{SL99}
Sabbagh AB, Lesser AJ.
J Polym Sci B: Polym Phys 1999; 37: 2651--2663.

\bibitem{MP01}
Meyer RW, Pruitt LA.
Polymer 2001; 42: 5293--5306.

\bibitem{BKR02}
Bergstr\"{o}m JS, Kurtz SM, Rimnac CM, Edidin AA.
Biomaterials 2002; 23: 2329--2343.

\bibitem{Seg02}
Seguela R.
J Polym Sci B: Polym Phys 2002; 40: 593--601.

\bibitem{SCC02}
Sweeney J, Collins TLD, Coates PD, Unwin AP, Duckett RA, Ward IM.
Int J Plasticity 2002; 18: 399--414.

\bibitem{AGU95}
Aboulfaraj M, G'Sell C, Ulrich B, Dahoun A.
Polymer 1995; 36: 731--742.

\bibitem{AFP95a}
Alberola N, Fugier M, Petit D, Fillon B.
J Mater Sci 1995; 30: 860--868.

\bibitem{AFP95b}
Alberola N, Fugier M, Petit D, Fillon B.
J Mater Sci 1995; 30: 1187--1195.

\bibitem{KY95}
O'Kane WJ, Young RJ.
J Mater Sci Lett 1995; 14: 433--435.

\bibitem{KYR95}
O'Kane WJ, Young RJ, Ryan AJ.
J Macromol Sci B Phys 1995; 34: 427--458.

\bibitem{SCC97}
Sweeney J, Collins TLD, Coates PD, Ward IM.
Polymer 1997; 38: 5991--5999.

\bibitem{CCG98}
Coulon G, Castelein G, G'Sell C.
Polymer 1998; 40: 95--110.

\bibitem{SCC99}
Sweeney J, Collins TLD, Coates PD, Duckett RA.
J Appl Polym Sci 1999; 72: 563--575.

\bibitem{SSE99}
Seguela R, Staniek E, Escaig B, Fillon B.
J Appl Polym Sci 1999; 71: 1873--1885.

\bibitem{NT99}
Nitta K-H, Takayanagi M.
J Polym Sci B: Polym Phys 1999; 37: 357--368.

\bibitem{NT00}
Nitta K-H, Takayanagi M.
J Polym Sci B: Polym Phys 2000; 38: 1037--1044.

\bibitem{RZF01}
Ran S, Zong X, Fang D, Hsia BS, Chu B, Phillips RA.
Macromolecules 2001; 34: 2569--2578.

\bibitem{LVS02}
Lima MFS, Vasconcellos MAZ, Samios D.
J Polym Sci B: Polym Phys 2002; 40: 896--903.

\bibitem{BJJ96}
Buckley CP, Jones DP, Jones DC.
Polymer 1996; 37: 2403--2414.

\bibitem{AGC96}
Ajji A, Guevremont J, Cole KC, Dumoulin MM, 
Polymer 1996; 37: 3707-3714.

\bibitem{MDW97}
Matthews RG, Duckett RA, Ward IM, Jones DP.
Polymer 1997; 38: 4795--4802.

\bibitem{ZB97}
Zaroulis JS, Boyce MC.
Polymer 1997; 38: 1303--1315.

\bibitem{ABJ98}
Adams AM, Buckley CP, Jones DP.
Polymer 1998; 39: 5761--5763.

\bibitem{LB99}
Llana PG, Boyce MC. 
Polymer 1999; 40: 6729--6751.

\bibitem{SNK99}
Suzuki A, Nakamura Y, Kunugi T.
J Polym Sci B: Polym Phys 1999; 37: 1703--1713.

\bibitem{BSL00}
Boyce MC, Socrate S, Llana PG. 
Polymer 2000; 41: 2183--2201.

\bibitem{KB97}
Kalay G, Bevis MJ.
J Polym Sci B: Polym Phys 1997; 35: 241--263, 265--291.

\bibitem{YMT98}
Yamada K, Matsumoto S, Tagashira K, Hikosaka M.
Polymer 1998; 39: 5327--5333.

\bibitem{ABM99}
Alamo RG, Brown GM, Mandelkern L, Lehtinen A, Paukkeri R.
Polymer 1999; 40: 3933--3944.

\bibitem{MHY00}
Maiti P, Hikosaka M, Yamada K, Toda A, Gu F.
Macromolecules 2000; 33: 9069--9075.

\bibitem{IS00}
Iijima M, Strobl G.
Macromolecules 2000; 33: 5204--5214.

\bibitem{GHT02}
Gu F, Hikosaka M, Toda A, Ghosh SK, Yamazaki S, Araki M, Yamada K.
Polymer 2002; 43: 1473--1481.

\bibitem{MF97}
Meddad A, Fisa B.
J Appl Polym Sci 1997; 64: 653--665.

\bibitem{DC02a}
Drozdov AD, Christiansen JdeC.
Polymer 2002; 43: 4745--4761.

\bibitem{QPR97}
Quinson R, Perez J, Rink M, Pavan A.
J Mater Sci 1997; 32: 1371--1379.

\bibitem{Hau00}
Haupt P.
Continuum mechanics and theory of materials,
Berlin: Springer, 2000.

\bibitem{Klo02}
Kloczkowski A.
Polymer 2002; 43: 1503--1525.

\bibitem{DC02b}
Drozdov AD, Christiansen JdeC.
Eur Polym J 2002; 38: in press.

\bibitem{KK74}
Kaye A, Kennett AJ.
Rheol Acta 1974; 13: 916--923.

\bibitem{HB95}
Hasan OA, Boyce MC. 
Polym Eng Sci 1995; 35: 331--344.

\bibitem{BK95}
Bordonaro CM, Krempl E.
Polym Eng Sci 1995; 35: 310--316.

\bibitem{BSG97}
Bardenhagen SG, Stout MG, Gray GT.
Mech Mater 1997; 25: 235--253.

\bibitem{SK98}
Spathis G, Kontou E.
Polymer 1998; 39: 135--142.

\bibitem{FB01}
Frank GJ, Brockman RA.
Int J Solids Structures 2001; 38: 5149--5164.

\bibitem{JTH02}
Jolly L, Tidu A, Heizmann I-J, Bolle B.
Polymer 2002; 43: in press.
\end{thebibliography}
\end{document}